\documentclass{svjour3}                     

\smartqed  
\usepackage{graphicx}
\usepackage{subcaption}

\usepackage{amsmath}  
\usepackage{amsfonts}
\usepackage{amssymb}
\usepackage{graphicx} 
\usepackage{graphics}
\usepackage{xspace}
\usepackage{color}
\usepackage{url}

\usepackage{cite}

\usepackage[normalem]{ulem}
\usepackage{color}

\newcommand{\dx}[1]{\hspace{-0.4em}\ensuremath{\mathrm{d}#1}\,}

\def\etab{\eta_{b}}

\def\K0bar{\overline{K^0}}
\def\bge{\begin{equation}}
\def\ene{\end{equation}}
\def\bg{\begin{eqnarray}}
\def\en{\end{eqnarray}}

\def\d0bar{{\bar{D}^0}}

\def\tbf{\textbf}

\newcommand{\be}{\begin{equation}}
\newcommand{\ee}{\end{equation}}
\newcommand{\bea}{\begin{eqnarray}}
\newcommand{\eea}{\end{eqnarray}}

\begin{document}

\title{
\boldmath{$\Upsilon$} and \boldmath{$\eta_b$} mass shifts in nuclear matter
}


\author{G.~N.~Zeminiani \and
        J.J.~Cobos-Mart\'{\i}nez \and
        K.~Tsushima}
        
\institute{G.~N.~Zeminiani  
\at Laborat\'orio de F\'isica Te\'orica e Computacional, 
Universidade Cidade de S\~ao Paulo (UNICID), 
01506-000, S\~ao Paulo, SP, Brazil\\
\email{guilherme.zeminiani@gmail.com}
\and 
J.J.~Cobos-Mart\'{\i}nez  
\at Departamento de F\'isica, Universidad de Sonora, Boulevard
Luis Encinas J. y Rosales, Colonia Centro, Hermosillo, Sonora 83000, M\'exico\\
\email{j.j.cobos.martinez@gmail.com, jesus.cobos@fisica.uson.mx}
\and 
K.~Tsushima 
\at 
Laborat\'orio de F\'isica Te\'orica e Computacional, 
Universidade Cidade de S\~ao Paulo (UNICID) and\\ 
Universidade Cruzeiro do Sul,  
01506-000, S\~ao Paulo, SP, Brazil\\
\email{kazuo.tsushima@gmail.com, kazuo.tsushima@cruzirodosul.edu.br}
}





\maketitle


\begin{abstract}
By extending the previous works that two of the present authors were 
involved, we estimate for the first time the $\Upsilon$ and $\eta_b$ as well as $B^*$ meson  
mass shifts (scalar potentials) in symmetric nuclear matter.
The main interest is, whether the strengths of the 
bottomonium-(nuclear matter) and charmonium-(nuclear matter)  
interactions are similar or very different, in the range of a few tens of MeV 
at the nuclear matter saturation density. This is because, each ($J/\Psi,\Upsilon$) and 
($\eta_c,\eta_b$) meson group is usually assumed to have very similar properties 
based on the heavy charm and bottom quark masses.
The estimate for the $\Upsilon$ is made using an SU(5) effective Lagrangian density 
and the anomalous coupling one, 
by studying the $BB$, $BB^*$, and $B^*B^*$ meson loop contributions  
for the self-energy in free space and in nuclear medium. 
As a result, we include only the $BB$ meson loop contribution 
as our prediction.
As for the $\eta_b$, to be complete, we include the $BB^*$ and $B^*B^*$ meson loop contributions   
in the self-energy for the analysis.
The in-medium masses of the $B$ and $B^{*}$ mesons appearing in the  
self-energy loops are calculated by the quark-meson coupling model. 
Form factors are used to regularize the loop integrals with 
a wide range of the cutoff mass values. 
A detailed analysis on the $BB$, $BB^{*}$, and $B^{*}B^{*}$ meson 
loop contributions for the $\Upsilon$ mass shift is made by comparing with
the respectively corresponding $DD, DD^*$, and $D^*D^*$ meson loop contributions 
for the $J/\Psi$ mass shift. Based on the analysis for the $\Upsilon$, 
our prediction for the $\eta_b$ mass shift is made on the same footing 
as that for the $\Upsilon$, namely including only the lowest order 
$BB^*$ meson loop. 
The $\Upsilon$ mass shift is predicted to be -16 to -22 MeV at the nuclear 
matter saturation density with the cutoff mass values in the range of 2000 - 6000 MeV  
using the $\Upsilon BB$ coupling constant determined by the vector 
meson dominance model with the experimental data, 
while the $\eta_b$ mass shift is predicted to be -75 to -82 MeV with 
the SU(5) universal coupling constant determined by the $\Upsilon BB$ 
coupling constant for the same range of the cutoff mass values.
Our results show an appreciable difference between the 
bottomonium-(nuclear matter) and charmonium-(nuclear matter) interaction strengths. 
We also study the $\Upsilon$ and $\eta_b$ mass shifts in a 
heavy quark (heavy meson) symmetry limit, namely, by calculating 
their mass shifts using the same coupling constant value as that was used 
to estimate the $J/\Psi$ and $\eta_c$ mass shifts. 
For the $\eta_b$ mass shift an SU(5) symmetry breaking case is also studied in this limit. 
Our predictions for these cases at nuclear matter saturation density are, 
-6 to -9 MeV for $\Upsilon$, -31 to -38 MeV for $\eta_b$, 
and -8 to -11 MeV for $\eta_b$ with a broken SU(5) symmetry, 
where the corresponding charm sector ones are, 
-5 to -21 for $J/\Psi$, -49 to -87 for $\eta_c$, and 
-17 to -51 for $\eta_c$ with a broken SU(4) symmetry.
\end{abstract}

\section{Introduction}
\label{intro}

The 12 GeV upgrade of CEBAF at the Jefferson Lab made it possible to produce low-momentum
heavy-quarkonia in an atomic nucleus. In a recent experiment~\cite{Ali:2019lzf}, a photon beam was 
used to produce a $J/\Psi$ meson near-threshold, which was identified by the decay into an 
electron-positron pair. Also with the construction of the FAIR facility in Germany, heavy and 
heavy-light mesons will be produced copiously by the annihilation of antiprotons 
on nuclei~\cite{Durante:2019hzd}.  

The production of heavy quarkonium in nuclei is one of the most useful methods for studying the
interaction of the heavy quarkonium with nucleon, in particular,
for probing its gluonic properties. 
We can, thus advance in understanding the hadron properties and their interactions based on 
quantum chromodynamics (QCD).  
Since the heavy quarkonium interacts with nucleon primarily via gluons, 
its production in a nuclear medium can be of great relevance 
to explore the roles of gluons. 

In the past few decades, many attempts 
were made~\cite{Hosaka:2016ypm,Krein:2016fqh,Metag:2017yuh,Krein:2017usp}  
to find alternatives to the meson-exchange mechanism for 
the (heavy-quarkonium)-nucleon interaction.
Some works employed charmed meson 
loops~\cite{Ko:2000jx,Krein:2010vp,Tsushima:2011kh,Tsushima:2011fg,Krein:2013rha}, 
others were based on QCD sum 
rules~\cite{Klingl:1998sr,Hayashigaki:1998ey,Kim:2000kj,Kumar:2010hs},  
phenomenological potentials~\cite{Belyaev:2006vn,Yokota:2013sfa}, 
the charmonium color polarizability~\cite{Peskin:1979va,Kharzeev:1995ij}, 
and van der Waals type 
forces~\cite{Kaidalov:1992hd,Luke:1992tm,deTeramond:1997ny,Brodsky:1997gh,Ko:2000jx,
Sibirtsev:2005ex,Voloshin:2007dx,TarrusCastella:2018php}.

Furthermore, lattice QCD simulations for charmonium-nucleon 
interaction in free space were performed in the last 
decade~\cite{Yokokawa:2006td,Liu:2008rza,Kawanai:2010ev,Kawanai:2010ru,Skerbis:2018lew}.
More recently, studies for the binding of charmonia with 
nuclear matter and finite nuclei, 
as well as light mesons and baryons, were performed in lattice QCD 
simulations~\cite{Beane:2014sda,Alberti:2016dru}. 
These simulations, however, used unphysically heavy pion masses.

In addition, medium modifications of charmed and bottom hadrons were 
studied in 
Refs.~\cite{Sibirtsev:1999js,Sibirtsev:1999jr,Tsushima:1998ru,Tsushima:2002cc,Tsushima:2002sm,
Tsushima:2003dd} based on the quark-meson coupling (QMC) model~\cite{Guichon:1987jp}, 
on which we will partly rely in this study. 
For example, based on the $D$ and $D^{*}$ meson mass modifications 
in symmetric nuclear matter calculated by the QMC model, 
the mass shift of $J/\Psi$ meson was predicted to be -16 to -24 MeV~\cite{Krein:2010vp} 
at the symmetric nuclear matter saturation density ($\rho_0 = 0.15$ fm$^{-3}$).
However, because of the unexpected contribution 
from the heavier $D^*D^*$ meson loop  
for the $J/\Psi$ self-energy, the authors updated the prediction for the 
$J/\Psi$ mass shift by including only the $DD$ meson loop~\cite{Krein:2017usp}. 
This gives the prediction of -3.0 to -6.5 MeV downward shift of the $J/\Psi$ mass at $\rho_0$.
In the QMC model the internal structure of hadrons changes in medium    
by the strong nuclear mean fields directly interacting with 
the light quarks $u$ and $d$, the present case in $D$ and $D^*$ mesons, 
and the dropping of these meson masses enhances  
the self-energy of $J/\Psi$ more than that in free space, 
resulting in an attractive $J/\Psi$-nucleus potential  
(negative mass shift)~\cite{Tsushima:2011kh}.

As for the $\eta_c$ meson, experimental studies of the production 
in heavy ion collisions at the LHC were  
performed~\cite{Aaij:2019gsn,Tichouk:2020dut,Tichouk:2020zhh,Goncalves:2018yxc,Klein:2018ypk}.
However, nearly no experiments were aimed to produce the $\eta_c$ 
at lower energies, probably due to the difficulties to perform experiment.
Furthermore, only recently the in-medium properties of $\eta_c$ meson were renewed 
theoretically~\cite{Cobos-Martinez:2020ynh}.

When it comes to the bottomonium sector on which we focus here, 
studies were made for $\Upsilon$ photoproduction at EIC 
(Electron-Ion Collider)~\cite{Xu:2020uaa,Gryniuk:2020mlh},  
$\Upsilon$ production in $p$Pb collisions~\cite{Aaij:2018scz}, 
and $\Upsilon(nl)$ (excited state) decay into $ B^{(*)} \bar B^{(*)}$~\cite{Liang:2019geg}. 
By such studies, we can further improve our understanding of the heavy 
quarkonium properties.
QCD predicts that chiral symmetry would be partially restored in a nuclear medium, and 
the effect of the restoration is expected to change the properties of hadrons in medium, 
particularly those hadrons that contain nonzero light quarks $u$ and $d$, 
because the reductions of the light quark $u$ and $d$ condensates are expected to be 
faster than those of the heavier quarks as nuclear density increases.  
Thus, usually the light quark condensates are regarded as the order parameters of the 
(dynamical) chiral symmetry. 
(Some studies support the faster reduction of the light-quark condensates in medium as nuclear 
density increases than those of the heavier quarks: (i) based on the NJL 
model~\cite{Tsushima:1991fe,Maruyama:1992ab} 
for the light and strange quark condensates in nuclear matter, 
and (ii) the result that the heavy quark condensates are proportional 
to the gluon condensate obtained by 
the operator product expansion~\cite{Shifman:1978bx} and also by a world-line effective 
action-based study~\cite{Antonov:2012ud}, together with the result of the model independent 
estimate that the gluon condensate at nuclear matter saturation density decreases only 
about 5\% by the QCD trace anomaly and Hellman-Feynman theorem~\cite{Cohen:1991nk}.) 

The frequently considered interactions between the heavy quarkonium and 
the nuclear medium are QCD van der Waals type 
interactions~\cite{Kaidalov:1992hd,Luke:1992tm, 
deTeramond:1997ny,Brodsky:1997gh,Ko:2000jx,Sibirtsev:2005ex, Voloshin:2007dx, 
TarrusCastella:2018php}. Naively, this must occur by the exchange of gluons 
in the lowest order, since heavy quarkonium has no light quarks,   
whereas the nuclear medium is composed of light quarks, and thus the light-quark or light-flavored 
hadron exchanges do not occur in this order. 
Another possible mechanism for the heavy quarkonium interaction with the nuclear medium 
is through the excitation of the intermediate state hadrons which contain light quarks. 

One of the simple, but interesting questions may be, 
whether or not the strengths of the charmonium-(nuclear matter)  
and bottomonium-(nuclear matter) interactions are indeed similar, 
since one often expects the similar properties of charmonium and bottomonium  
based on the heavy charm and bottom quark masses. 

\begin{figure}[htb]
\centering 
  \includegraphics[scale=0.9]{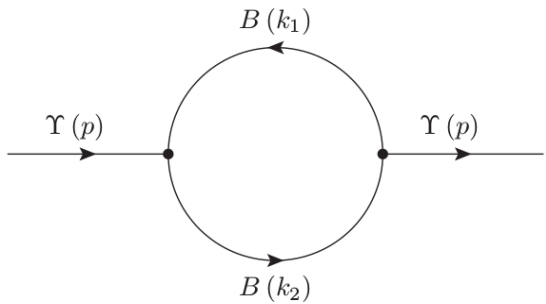}
 \caption{BB meson loop contribution for the $\Upsilon$ self-energy.}
 \label{fig1}
\end{figure}

In this article, after calculating the in-medium $B$ and $B^*$ meson masses, 
we estimate first the mass shift of $\Upsilon$ meson in terms of 
the excitations of intermediate state hadrons with light quarks  
in the self-energy. As an example we show in 
Fig.~\ref{fig1} the $BB$ meson loop contribution for 
the $\Upsilon$ self-energy --- we will also study the $BB^*$ and $B^*B^*$ meson loop contributions.
Next, we also estimate the mass shift of the pseudoscalar 
quarkonium, $\eta_b$ meson, which is the lightest $b\bar{b}$ bound state. 
The estimates will be made using an SU(5) effective 
Lagrangian density (hereafter we will denote simply by ''Lagrangian'') 
which contains both the $\Upsilon$ and $\eta_b$ mesons with one universal coupling constant, 
and the anomalous coupling one respecting an SU(5) 
symmetry in the coupling constant.
Then, the present study can also provide information 
on the SU(5) symmetry breaking. 

Upon expanding the SU(5) effective Lagrangian with minimal substitutions, 
we get the interaction Lagrangians for calculating the $\Upsilon$  
self-energy, i.e., the $BB$, and $B^*B^*$ meson loops   
as well as for the the $\eta_b$ self-energy, the $BB^*$ meson loop 
($\Upsilon BB^*$ and $\eta_bB^*B^*$ interaction Lagrangians are  
anomalous coupling ones,  
not derived from the SU(5) effective Lagrangian). 
Thus, we need to have better knowledge on the in-medium properties (Lorentz-scalar  
and Lorentz-vector potentials) of the $B$ and  $B^{*}$ mesons.
For this purpose we use the QMC model invented by Guichon~\cite{Guichon:1987jp},  
which has been successfully applied for various 
studies~\cite{Krein:2017usp,Krein:2010vp,Tsushima:2002cc,Tsushima:1997df,Guichon:1995ue, 
Saito:1996sf,Tsushima:2019wmq,Saito:2005rv}.
The QMC model is a phenomenological, but very successful quark-based relativistic mean 
field model for nuclear matter, nuclear structure, and hadron properties 
in a nuclear medium. 
The model relates the relativistically moving  
confined light $u$ and $d$ quarks in the nucleon bags with the scalar-isoscalar ($\sigma$), 
vector-isoscalar ($\omega$), and vector-isovector ($\rho$) mean fields self-consistently 
generated by the light quarks in the nucleons~\cite{Guichon:1987jp}.
Note that, the in-medium $B^*$ meson mass is estimated and presented for the first 
time in this study, calculated by the QMC model. 

We analyze the $BB, BB^*$ and $B^*B^*$ meson loop contributions    
for the $\Upsilon$ self-energy. After a detailed analysis, 
our predictions for the $\Upsilon$ and $\eta_b$ mass shifts
are made by including only the lowest order $BB$ meson loop contribution for 
the $\Upsilon$, 
and only the $BB^*$ meson loop contribution for the $\eta_b$,  
where the in-medium masses of the $B$ and $B^*$ mesons are calculated by the QMC model.
In addition, a detailed comparison is made between the $\Upsilon$ and $J/\Psi$ meson self-energies, 
in order to get a better insight into the cutoff mass values used in the form factors, 
as well as the form factors themselves.

This article is organized as follows. 
In Sec.~\ref{qmc} we describe the $B$ and $B^*$ mesons 
in symmetric nuclear matter within the QMC model. 
We present in Sec.~\ref{umass} the effective Lagrangians obtained from  
a unified SU(5) symmetry Lagrangian by minimal substitutions, 
as well as the anomalous coupling one, and first study the $\Upsilon$ 
self-energy. 
The results for the $BB$, $BB^*$ and $B^*B^*$ meson loop contributions  
for the $\Upsilon$ self-energy are studied in detail, 
and the case of the total ($BB+BB^*+B^*B^*$) meson loop contribution 
and its decomposition are compared  
with the corresponding total ($DD+DD^*+D^*D^*$) meson loop contribution 
and its decomposition in the $J/\Psi$ self-energy.
In Sec.~\ref{emass} we study the $\eta_b$ mass shift 
including the $BB^*$ and $B^*B^*$ meson loop contributions, and present our  
prediction by taking only the $BB^*$ meson loop contribution, to be consistent 
with the prediction for the $\Upsilon$ mass shift. 
In Sec.~\ref{hlimit} we consider a heavy quark (heavy meson) symmetry limit 
for the $\Upsilon$ and $\eta_b$, and a broken SU(5) symmetry for the $\eta_b$ 
in this limit, and also give predictions for these cases. 
We perform in Sec.~\ref{diff_ff} an initial study for the effects of the form factor  
on the $\Upsilon$ and $\eta_b$ mass shifts using a different form factor. 
Lastly, summary and conclusion are given in Sec.~\ref{con}.

\section{Quark-meson coupling model}
\label{qmc}

In this section we focus on the properties of $B$ and $B^*$ mesons 
in symmetric nuclear matter, and calculate their Lorentz-scalar effective masses using the 
QMC model, where the in-medium $B^*$ meson mass has not been calculated nor presented in the past. 
This is enough, since the vector potentials cancel out   
in each $BB, BB^*$, and $B^*B^*$ meson loop calculation for the $\Upsilon$ and $\eta_b$  
self-energies, namely, they cancel out in the energy-contour integral in each meson loop, 
and this is consistent with the baryon number conservation at the quark level, 
since the vector mean filed potential proportionals to baryon density.

The QMC model is a quark-based model for nuclear matter and
finite nuclei by describing the internal structure of the nucleon  
using the MIT bag (original version~\cite{Guichon:1987jp}), 
and the binding of nucleons by the self-consistent 
couplings of the confined light quarks $u$ and $d$ to the scalar-$\sigma$ 
and vector-isoscalar-$\omega$ and vector-isovector-$\rho$ 
meson fields generated by the confined light quarks
in the nucleons~\cite{Guichon:1987jp,Guichon:1995ue,Saito:1996sf}.
In a nuclear medium, the hadrons with light quarks 
are expected to change their properties predominantly, and thus affect 
the interaction with nucleons, 
what makes the QMC model a useful model to describe the change of the internal structure of 
hadrons in a nuclear medium.

Assuming SU(2) symmetry for the quarks ($m_q = m_u = m_d$ and $q = u$ or $d$ below) 
as well as for nucleons, the Dirac equations for the quarks and antiquarks in
nuclear matter, inside the bags of $B$ and $B^{*}$ mesons  
embedded in nuclear matter neglecting the 
Coulomb force, are given by~\cite{Tsushima:1997df,Tsushima:2002cc}:
\begin{eqnarray}
&&\left[i\gamma \cdot \partial_{x} - \left(m_{q} - V^{q}_{\sigma}\right)
\mp \gamma^{0} \left(V^{q}_{\omega} + \frac{1}{2}V^{q}_{\rho}\right)\right]
\begin{pmatrix}
        \psi_{u}\left(x\right)\\
        \psi_{\overline{u}}\left(x\right)
       \end{pmatrix} = 0,\\
&&\left[i\gamma \cdot \partial_{x} - \left(m_{q} - V^{q}_{\sigma}\right)
\mp \gamma^{0} \left(V^{q}_{\omega} - \frac{1}{2}V^{q}_{\rho}\right)\right]
\begin{pmatrix}
        \psi_{d}\left(x\right)\\
        \psi_{\overline{d}}\left(x\right)
       \end{pmatrix} = 0,\\
&&\left[i\gamma \cdot \partial_{x} - m_{b}\right]\psi_{b, \overline{b}}\left(x\right) = 0.
\end{eqnarray}
In the above, the (constant) mean-field potentials for the light quark $q$ in nuclear
matter are defined by $V^{q}_{\sigma} \equiv g^{q}_{\sigma}\sigma$, 
$V^{q}_{\omega} \equiv g^{q}_{\omega}\omega = g^q_\omega\, \delta^{\mu,0} \omega^\mu$, 
$V^{q}_{\rho} \equiv g^{q}_{\rho}b = g^q_\rho\, \delta^{i,3} \delta^{\mu,0} \rho^{i,\mu}$, 
with the $g^{q}_{\sigma}$, $g^{q}_{\omega}$ and
$g^{q}_{\rho}$ being the corresponding quark-meson coupling constants.

The static solution for the ground state quarks (antiquarks) with a flavor
$f (=u,d,b)$ is written as 
$\psi_{f}\left(x\right) = N_{f}e^{-i\epsilon_{f}t/R^{*}_{B,B^{*}}}\psi_{f}\left(\textbf{r}\right)$, 
with the normalization factor $N_{f}$, and $\psi_{f}\left(\textbf{r}\right)$ the corresponding 
spin and spatial part of the wave function.
The eigenenergies for the quarks and antiquarks in the $B$ and $B^{*}$ mesons in units of 
$1/R^{*}_{B,B^{*}}$ are given by:
\begin{eqnarray}
&&\begin{pmatrix}
        \epsilon_{u}\left(x\right)\\
        \epsilon_{\overline{u}}\left(x\right)
       \end{pmatrix} = \Omega^{*}_{q} \pm R^{*}_{B,B^{*}} 
\left(V^{q}_{\omega} + \frac{1}{2}V^{q}_{\rho}\right),\\
&&\begin{pmatrix}
        \epsilon_{d}\left(x\right)\\
        \epsilon_{\overline{d}}\left(x\right)
       \end{pmatrix} = \Omega^{*}_{q} \pm R^{*}_{B,B^{*}} 
\left(V^{q}_{\omega} - \frac{1}{2}V^{q}_{\rho}\right),\\
&&\epsilon_{b} = \epsilon_{\overline{b}} = \Omega_{b}.
\end{eqnarray}

\begin{figure}[htb]
\vspace{4ex}
\centering
 \includegraphics[scale=0.35]{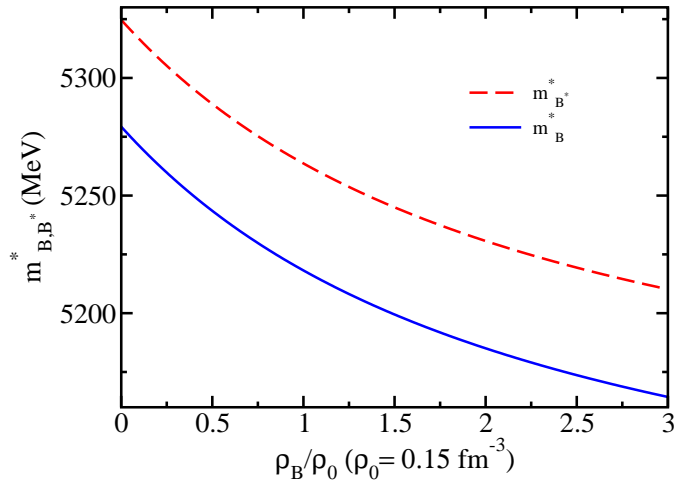}
 \caption{$B$ and $B^{*}$ meson effective Lorentz-scalar masses  
 in symmetric nuclear matter versus baryon density.}
 \label{fig2}
\end{figure}

The $B$ and $B^{*}$ meson masses in a nuclear medium, $m^{*}_{B, B^{*}}$, are calculated by
\begin{eqnarray}
&& m^{*}_{B,B^{*}} = \sum_{j=q,\overline{q},b,\overline{b}}
\frac{n_{j}\Omega^{*}_{j}- Z_{B,B^{*}}}{R^{*}_{B,B^{*}}} + \frac{4}{3} \pi R^{*3}_{B,B^{*}}B_{p},
\\
&&\left. \frac{d m^{*}_{B,B^{*}}}{d R_{B,B^{*}}}\right|_{R_{B,B^{*}} = R^{*}_{B,B^{*}}} = 0, 
\end{eqnarray}
where $\Omega^{*}_{q} = \Omega^{*}_{\overline{q}} = \left[x^{2}_{q} + \left(R^{*}_{B,B^{*}} 
m^{*}_{q}\right)^{2}\right]^{1/2}$, and $m^{*}_{q} = m_{q} - g^{q}_{\sigma}\sigma$
and $\Omega^{*}_{b} = \Omega^{*}_{\overline{b}} = \left[x^{2}_{b} + \left(R^{*}_{B,B^{*}} 
m_{b}\right)^{2}\right]^{1/2}$, with $x_{q,b}$ being the lowest mode bag eigenfrequencies. 
$B_{p}$ is the bag constant, and $n_{q,b}$ ($n_{\overline{q},\overline{b}}$) are the lowest 
mode valence quark (antiquark) numbers 
for the quark flavors $q$ and $b$ 
in the corresponding $B$ and $B^{*}$ mesons, 
and the $z_{B,B^{*}}$ parameterize the sum of the center-of-mass 
and gluon fluctuation effects and are 
assumed to be independent of density~\cite{Guichon:1995ue}.

We choose the values ($m_q, m_b$) = (5, 4200) MeV for the current quark masses, and $R_{N}$ 
= 0.8 fm for the free space nucleon bag radius. 
The quark-meson coupling constants, 
$g^{q}_{\sigma}$, $g^{q}_{\omega}$ and $g^{q}_{\rho}$ used for the light quarks   
in the $B$ and $B^{*}$ mesons (the same as in the nucleon), were determined by the fit 
to the saturation energy (-15.7 MeV) at the saturation density  
($\rho_0 = 0.15$ fm$^{-3}$) of symmetric nuclear matter for $g^q_\sigma$ and $g^q_\omega$, 
and by the bulk symmetry energy (35 MeV) for $g^q_\rho$~\cite{Guichon:1987jp,Saito:2005rv}.
The QMC model predicts a similar amount in the decrease of the in-medium effective 
Lorentz-scalar masses of the $B$ and $B^{*}$ mesons in symmetric nuclear matter 
as shown in Fig.~\ref{fig2}.
At $\rho_0$ the mass shifts of the $B$ and $B^*$ mesons are respectively, 
$(m^*_B - m_B)=-61$ MeV and $(m^*_{B^*}-m_{B^*})=-61$ MeV, the difference 
in their mass shift values appears in the next digit.
To calculate the $\Upsilon$ and $\eta_b$ meson self-energies in symmetric nuclear matter 
by the excited $B$ and $B^*$ meson intermediate states in the loops, 
we use the calculated in-medium masses of them shown in Fig.~\ref{fig2}.
Note that, when we use ($m_q, m_b$) = (5, 4180) and (5, 4200) MeV, 
the obtained in-medium masses are,  
($m_B^*, m_{B^*}^*$) = (5218.207, 5263.687) and (5218.170, 5263.652) MeV at $\rho_0$  
[(5164.554, 5210.326) and (5164.385, 5210.162) at $3\rho_0$], respectively. 
Thus, the $m_b$ value dependence is negligible for the present purpose.
\\
(The same argument also holds for the charm quark mass $m_c$ value dependence
of the in-medium masses, 
$m_D^*$ and $m_{D^*}^*$ used for the $J/\psi$ and $\eta_c$ mass shift. 
In this case, for ($m_q, m_c$) = (5, 1270) and (5, 1300) MeV,  
($m_D^*, m_{D^*}^*$) = (1805.235, 1946.943) and (1805.232, 1946.940) MeV 
at $\rho_0$ [(1748.380, 1891.227) and (1748.376, 1891.223) at $3\rho_0$], respectively.)

\section{$\Upsilon$ mass shift}
\label{umass}

\subsection{Effective Lagrangians and $\Upsilon$ self-energy}
\label{uslef}

The $\Upsilon$ mass shift in medium comes from the modifications of the $BB$, $BB^{*}$ and 
$B^{*}B^{*}$ meson loop contributions to the $\Upsilon$ self-energy relative to those in free space,
where the self-energy is calculated based on an effective flavor SU(5) symmetry 
Lagrangian~\cite{Lin:2000ke}, 
and the anomalous coupling one, to be specified later.
The free Lagrangian for pseudoscalar and vector mesons is given by, 
\begin{equation}
{\cal L}_{0}=Tr \left(  \partial _{ \mu }P^{\dagger} \partial ^{ \mu }P \right)
-\frac{1}{2}Tr \left( F_{ \mu  \nu }^{\dagger}F^{ \mu  \nu } \right),
\end{equation}
with \[ F_{ \mu  \nu }= \partial _{ \mu }V_{ \nu }- \partial _{ \nu }V_{ \mu } ,\]
where $P$ and $V$ are, respectively, the $5 \times 5$ pseudoscalar 
and vector meson matrices in SU(5):
\begin{eqnarray}
&&\hspace{-4ex} P = \frac{1}{\sqrt{2}} \begin{pmatrix} 
\frac{\pi^{0}}{\sqrt{2}} + \frac{\eta}{\sqrt{6}} + \frac{\eta_{c}}{\sqrt{12}} 
+ \frac{\eta_{b}}{\sqrt{20}}  &  \pi^{+}  &  K^{+}  &  \overline{D}^{0}  &  B^{+}\\
\pi^{-}  &  \frac{-\pi^{0}}{\sqrt{2}} + \frac{\eta}{\sqrt{6}} + \frac{\eta_{c}}{\sqrt{12}} 
+ \frac{\eta_{b}}{\sqrt{20}}  &  K^{0}  &  D^{-}  &  B^{0}\\
K^{-}  &  \overline{K}^{0}  &  \frac{-2\eta}{\sqrt{6}} + \frac{\eta_{c}}{\sqrt{12}} 
+ \frac{\eta_{b}}{\sqrt{20}}  &  D_{s}^{-}  &  B_{s}^{0}\\
D^{0}  &  D^{+}  &  D_{s}^{+}  &  \frac{-3\eta_{c}}{\sqrt{12}} 
+ \frac{\eta_{b}}{\sqrt{20}}  &  B_{c}^{+}\\
B^{-}  &  \overline{B^{0}}  &  \overline{B_{s}^{0}}  &  B_{c}^{-}  &  \frac{-2\eta_{b}}{\sqrt{5}}\\
\end{pmatrix}, \label{p} \\
\nonumber\\
\nonumber\\
&&\hspace{-4ex} V = \frac{1}{\sqrt{2}} \begin{pmatrix}
\frac{\rho^{0}}{\sqrt{2}} + \frac{\omega}{\sqrt{6}} + \frac{J/\Psi}{\sqrt{12}} 
+ \frac{\Upsilon}{\sqrt{20}}  &  \rho^{+}  &  K^{*+}  &  \overline{D}^{*0}  &  B^{*+}\\
\rho^{-}  &  \frac{-\rho^{0}}{\sqrt{2}} + \frac{\omega}{\sqrt{6}} + \frac{J/\Psi}{\sqrt{12}} 
+ \frac{\Upsilon}{\sqrt{20}}  &  K^{*0}  &  D^{*-}  &  B^{*0}\\
K^{*-}  &  \overline{K}^{*0}  &  \frac{-2\omega}{\sqrt{6}} + \frac{J/\Psi}{\sqrt{12}} 
+ \frac{\Upsilon}{\sqrt{20}}  &  D_{s}^{*-}  &  B_{s}^{*0}\\
D^{*0}  &  D^{*+}  &  D_{s}^{*+}  &  \frac{-3J/\Psi}{\sqrt{12}} + \frac{\Upsilon}{\sqrt{20}}  &  
B_{c}^{*+}\\
B^{*-}  &  \overline{B^{*0}}  &  \overline{B_{s}^{*0}}  &  B_{c}^{*-}  &  
\frac{-2\Upsilon}{\sqrt{5}} \\ 
\end{pmatrix}. \label{v}
\end{eqnarray}

The following minimal substitutions are introduced to obtain the couplings (interactions) 
between the pseudoscalar mesons and vector mesons~\cite{Lin:2000ke}:
\begin{eqnarray}
&&\partial _{ \mu }P \rightarrow  \partial _{ \mu }P-\frac{ig}{2} \left[ V_{ \mu }\text{, P} 
\right],\\
&&F_{ \mu  \nu } \rightarrow  \partial _{ \mu }V_{ \nu }- \partial _{ \nu }V_{ \mu }
 -\frac{ig}{2} \left[ V_{ \mu },~V_{ \nu } \right].
\end{eqnarray}

Then, the effective Lagrangian is obtained as, 
\begin{eqnarray}
 {\cal L}&={\cal L}_{0}+igTr \left(  \partial _{ \mu }P \left[ P,~V_{ \mu } \right]  \right) 
 -\frac{g^{2}}{4}Tr \left(  \left[ \text{P, V}_{ \mu } \right] ^{2} \right)\nonumber \\ 
 &+igTr \left(  \partial ^{ \mu }V^{ \nu } \left[ V_{ \mu },~V_{ \nu } \right]  \right) 
 +\frac{g^{2}}{8}Tr \left(  \left[ V_{ \mu },~V_{ \nu } \right] ^{2} \right). 
\label{Lint}
\end{eqnarray}

Expanding this in terms of the components given in Eqs.~(\ref{p}) and~(\ref{v}), 
we obtain the following interaction Lagrangians~\cite{Lin:2000ke},  
\begin{eqnarray}
{\cal L}_{\Upsilon BB} 
&=& i g_{\Upsilon BB}\Upsilon^{\mu}
\left[\overline{B} \partial_{\mu}B 
 - \left(\partial_{\mu}\overline{B}\right)B\right],
\\
{\cal L}_{\Upsilon B^{*}B^{*}} 
&=& i g_{\Upsilon B^{*}B^{*}}
\left\{ \Upsilon^{\mu}\left[ (\partial_{\mu}\overline{B^{*}}^{\nu}) B^{*}_{\nu} - 
\overline{B^{*}}^{\nu}\partial_{\mu} B^{*}_{\nu}\right]
+ \left[ (\partial_{\mu} \Upsilon^{\nu}) \overline{B^{*}_{\nu}} - \Upsilon^{\nu}
\partial_{\mu} \overline{B^{*}_{\nu}}\right] 
B^{*\mu}
\right.
\\ 
&&\hspace{12ex} 
\left.+ \overline{B^{*}}^{\mu} \left[\Upsilon^{\nu} \partial_{\mu} B^{*}_{\nu} - 
(\partial_{\mu}\Upsilon^{\nu}) B^{*}_{\nu}\right]\right\},
\end{eqnarray}
where the following convention is adopted
\begin{align*}
B&=\begin{pmatrix}
        B^{+}\\
        B^{0}
       \end{pmatrix}, & \overline{B}=\begin{pmatrix}
       B^{-} & \overline{B^{0}} \end{pmatrix},  
			& &B^{*} =\begin{pmatrix}
        B^{*+}\\
        B^{*0}
       \end{pmatrix}, & &\overline{B^{*}}&=\begin{pmatrix}
       B^{*-} & \overline{B^{*0}} \end{pmatrix}.\\         
\end{align*}

We obtain the coupling constants by the vector meson dominance 
(VMD) hypothesis (model)~\cite{Sakurai:1960ju,Sakurai,Lin:2000ke} using the experimental data 
for $\Gamma(\Upsilon \to e^+ e^-)$, 
\begin{equation}
 g_{\Upsilon BB} = g_{\Upsilon B^{*}B^{*}} = \frac{5g}{4\sqrt{10}} = 13.2228 \simeq 13.2.
\end{equation}
Note that, the use of the same form of the effective Lagrangian, the VMD model, and 
the $\Gamma(J/\Psi \to e^+ e^-)$, $g_{\Psi DD} = 7.46$ was obtained and used 
in Ref.~\cite{Krein:2010vp}.
In obtaining $g_{\Upsilon BB} = 13.2$ ($g_{\Psi DD} = 7.46$) by the VMD model 
with the data for $\Upsilon \to e^+ e^-$ ($J/\Psi \to e^+ e^-$), 
the $b$ ($c$) quark charge $e_b$ ($e_c$) and vector meson mass $m_\Upsilon$ ($m_{\Psi}$) 
enter as $|g_{\Upsilon BB}| \propto |e_b| \sqrt{m_\Upsilon/\Gamma(\Upsilon \to e^+ e^-)}$ 
( $|g_{\Psi DD}| \propto |e_c| \sqrt{m_{\Psi}/\Gamma(J/\Psi \to e^+ e^-)}$ ), 
and this results in a large difference for the obtained coupling constants 
between $g_{\Upsilon BB}$ and $g_{\Psi DD}$, where we have suppressed the common constant factor.
(See Appendix A of Ref.~\cite{Lin:1999ad} for details.)
Thus, one can expect a large SU(5) breaking for the charm and bottom quark sector mass shifts.

In addition we also include the $\Upsilon BB^*$ 
anomalous-coupling~\cite{Eletsky:1982py,Leinweber:2001ac} interaction Lagrangian, 
similar to the case of $J/\Psi$ that was introduced in the $J/\Psi DD^*$ interaction 
Lagrangian in Refs.~\cite{Oh:2000qr,Krein:2010vp}, 
\be
{\cal L}_{\Upsilon BB^{*}}
= \frac{g_{\Upsilon BB^{*}}}{m_{\Upsilon}}\varepsilon_{\alpha \beta \mu \nu}
\left(\partial^{\alpha}\Upsilon^{\beta}\right)\left[\left(\partial^{\mu} 
\overline{B^{*}}^{\nu}\right)B
+ \overline{B}\left(\partial^{\mu}{B^{*}}^{\nu}\right)\right],
\ee
where, we assume $g_{\Upsilon BB^*} = g_{\Upsilon BB} = g_{\Upsilon B^*B^*}$,  
the corresponding relation adopted for the $J/\Psi$ case~\cite{Krein:2010vp}.

The in-medium potential for the $\Upsilon$ meson is the difference of the in-medium, 
$m^{*}_{\Upsilon}$, and free space, 
$m_{\Upsilon}$, masses of $\Upsilon$,
\begin{equation}
 V = m_{\Upsilon}^{*} - m_{\Upsilon},
\end{equation}
with the free space physical $\Upsilon$ mass being reproduced first by,   
\begin{equation}
m^{2}_{\Upsilon} = \left(m^{0}_{\Upsilon}\right)^{2} + 
\Sigma_\Upsilon (k^{2}=m^{2}_{\Upsilon}),
\end{equation}
where $m^{0}_{\Upsilon}$ is the bare mass, and the total self-energy 
$\Sigma_\Upsilon$ is calculated by the sum
of the contributions from the free space $BB$, $BB^{*}$ and $B^{*}B^{*}$ meson loops    
in the case we include all the meson loops considered in this study. 
Note that, we ignore the possible width, or the imaginary part in the self-energy in the present  
study. The in-medium mass, $m^{*}_{\Upsilon}$,
is calculated likewise, by the total self-energy in medium using the medium-modified $B$ and 
$B^{*}$ meson masses with the same $m_\Upsilon^0$ value fixed in free space.
We remind that the $m^0_\Upsilon$ value depends on the loops included in the self-energy 
of the $\Upsilon$ in free space.

We sum each meson loop contribution for the $\Upsilon$ self-energy as
\begin{equation}
\Sigma_{\Upsilon} = \sum_l \Sigma_\Upsilon^l = \sum_l (- \frac{g^{2}_{\Upsilon l}}{3\pi^{2}})
\int_{0}^{\infty} dq\, \textbf{q}^{2} F_{l}(\textbf{q}^{2}) 
K_{l}(\textbf{q}^{2}),
\end{equation}
where $l = BB, BB^{*}, B^{*}B^{*}$ and $F_{l}\left(\textbf{q}\right)$ is the product of 
vertex form factors (to be discussed later). The $K_{l}$ for each meson loop contribution 
is given, similarly to the $J/\Psi$ case~\cite{Krein:2010vp},
\begin{eqnarray}
&&K_{BB}\left(\textbf{q}^{2}\right) = \frac{1}{\omega_{B}} 
\left(\frac{\textbf{q}^{2}}{\omega^{2}_{B} - m^{2}_{\Upsilon}/4}\right),
\\
&&K_{BB^{*}}\left(\textbf{q}^{2}\right) =\frac{\textbf{q}^{2}\overline{\omega}_{B}} 
{\omega_{B}\omega_{B^{*}}} \frac{1}{\overline{\omega}^{2}_{B} - m^{2}_{\Upsilon}/4},
\\
&&K_{B^{*}B^{*}}\left(\textbf{q}^{2}\right) =\frac{1}{4m_{\Upsilon} \omega_{B^*} }
\left[
\frac{A\left(q^{0} = \omega_{B^*} \right)}{\omega_{B^*} - m_\Upsilon/2} 
- \frac{A\left( q^{0} = \omega_{B^*} + m_{\Upsilon} \right)}{\omega_{B^*} + m_{\Upsilon}/2}
\right],
\end{eqnarray}
where $\omega_{B} = \left(\textbf{q}^{2} + m^{2}_{B}\right)^{1/2}$,
$\omega_{B^{*}} = \left(\textbf{q}^{2} + m^{2}_{B^{*}}\right)^{1/2}$,
$\overline{\omega}_{B} = \left(\omega_{B} + \omega_{B^{*}}\right)$ and

\begin{equation}
A\left(q\right) = \sum^{4}_{i=1} A_{i}\left(q\right),
\end{equation}
with
%
\begin{eqnarray}
A_1 (q) 
&=& -4q^2 \left[
4-\frac{q^2 + (q-k)^2}{m^{2}_{B^*}}
+ \frac{\left[q \cdot (q-k) \right]^2}{m^{4}_{B^*}} 
\right],
\label{EqA1}
\\
A_2 (q) 
&=& 8\left[q^2 - \frac{\left[q \cdot (q-k) \right]^2}{m^2_{B^*}}\right]
\left[2 + \frac{(q^0)^{2}}{m^{2}_{B^{*}}} \right],
\\
A_3 (q ) 
&=& 8\, (2q^0 - m_\Upsilon)
\left[
q^0 - (2q^{0} - m_\Upsilon) \frac{q^2 + q \cdot (q-k)}{m^{2}_{B^*}}
+q^{0}\frac{\left[q \cdot\left(q-k\right)\right]^{2}}{m^{4}_{B^*}}
\right],
\\
A_4 (q) 
&=& -8 \left[ q^0 - (q^0 - m_\Upsilon) \frac{q\cdot(q-k)}{m^2_{B^*}} \right]  
\left[ (q^0 - m_\Upsilon) - q^0\frac{q\cdot(q-k)}{m^2_{B^*}} \right],	
\label{EqA4}
\end{eqnarray}
where $q = \left(q^{0}, \textbf{q}\right)$, and the $\Upsilon$ is taken at rest, 
$k = \left(m_{\Upsilon}, 0\right)$.\par

We use phenomenological form factors to regularize the self-energy
loop integrals following Refs.~\cite{Krein:2010vp,Leinweber:1999ig}, 
\begin{equation}
u_{B,B^{*}}(\textbf{q}^{2}) = \left(\frac{\Lambda^{2}_{B,B^{*}} + m^{2}_{\Upsilon}}
{\Lambda^{2}_{B,B^{*}} + 4\omega^{2}_{B,B^{*}}\left(\textbf{q}^{2}\right)}\right)^{2}.
\label{ffups}
\end{equation}

For the vertices $\Upsilon BB$, $\Upsilon BB^{*}$ and $\Upsilon B^{*}B^{*}$, we use the form 
factors 
$F_{BB}\left(\textbf{q}^{2}\right) = u^{2}_{B}\left(\textbf{q}^{2}\right)$, 
$F_{BB^{*}}\left(\textbf{q}^{2}\right) = 
u_{B}\left(\textbf{q}^{2}\right)u_{B^{*}}\left(\textbf{q}^{2}\right)$, and 
$F_{B^{*}B^{*}}\left(\textbf{q}^{2}\right) = u^{2}_{B^{*}}\left(\textbf{q}^{2}\right)$, 
respectively, with $\Lambda_B$ ($\Lambda_{B^*}$) being the corresponding cutoff 
mass associated with $B$ ($B^*$) meson, and the common value, $\Lambda_{B} = \Lambda_{B^{*}}$,
will be used in this study. 

We have to point out that the choice of the cutoff mass values in the form factors for the 
$\Upsilon BB$, $\Upsilon BB^{*}$ and $\Upsilon B^{*}B^{*}$ vertices 
has nonnegligible impact on the results.
But the form factors are necessary to include the effects of the finite 
sizes of the mesons for the overlapping regions associated with the vertices. 
The cutoff values $\Lambda_{B,B^*}$ may be associated with the energies used to probe 
the internal structure of the mesons or the overlapping regions associated with the vertices. 
When these values get closer to the corresponding meson masses, 
the Compton wavelengths associated with the values of $\Lambda_{B,B^*}$ 
are comparable to the sizes of the mesons, 
and the use of the form factors does not make reasonable sense.
Then, in order to have a physical meaning, we may be able to 
constrain the choice for the cutoff mass values,   
in such a way that the form factors reflect the finite size effect of the participating mesons 
reasonably. Later, an analysis on this issue will be made   
taking the $J/\Psi DD$, $J/\Psi DD^*$ and $J/\Psi D^*D^*$  
vertices as examples.

By the heavy quark and heavy meson symmetry in QCD, the charm and bottom quark sectors  
are expected to have mostly similar properties (but quantitatively need to be shown if possible). 
Then, we follow this naive expectation and choose the similar cutoff mass values 
as the ones used in the previous work of the $J/\Psi$ mass shift~\cite{Krein:2017usp}, 
varying the $\Lambda_{B,B^*}$ values   
between $2000~\text{MeV} \leq \Lambda_{B,B^*} \leq 6000~\text{MeV}$, 
but with the larger upper values, since the $B$ and $B^*$ masses are 
larger than those of the $D$ and $D^*$ mesons.

\subsection{Results for {$\Upsilon$} mass shift}
\label{ures}

In the following we present the results for the in-medium mass shift of $\Upsilon$ meson 
together with each meson loop contribution    
for five different values of the cutoff mass $\Lambda_B (= \Lambda_{B^*})$, 
where we use the in-medium $B$ and $B^{*}$ meson masses shown in Fig.~\ref{fig2}.
The values used for the free space masses of $\Upsilon$, $B$ and $B^{*}$ mesons are, 
respectively, 9460, 5279 and 5325 MeV~\cite{PDG2020}.

\begin{figure}[htb]%
\vspace{4ex}
\centering
\includegraphics[scale=0.34]{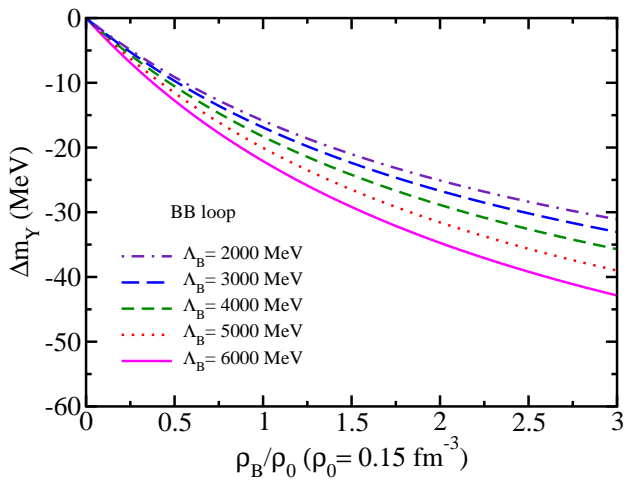}%
\caption{Contribution from the $BB$ meson loop (total is $BB$ meson loop) to 
the $\Upsilon$ mass shift versus nuclear matter density for five different values of 
the cutoff mass $\Lambda_{B}$.}%
\label{fig3}%
\end{figure}

\begin{figure}[htb]%
\vspace{2ex}
\centering
\includegraphics[width=6.5cm]{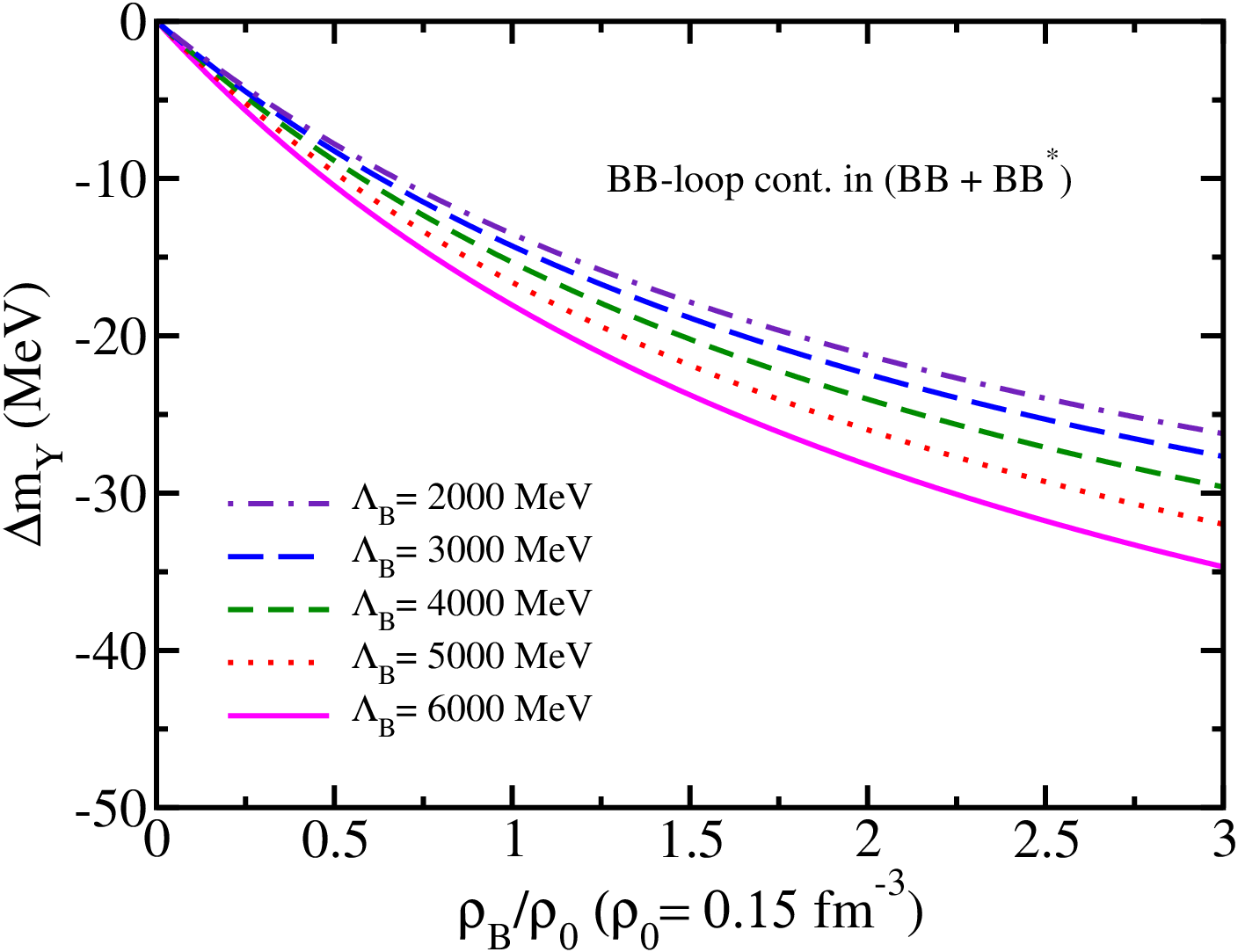}
\hspace{2ex}
\includegraphics[width=6.5cm]{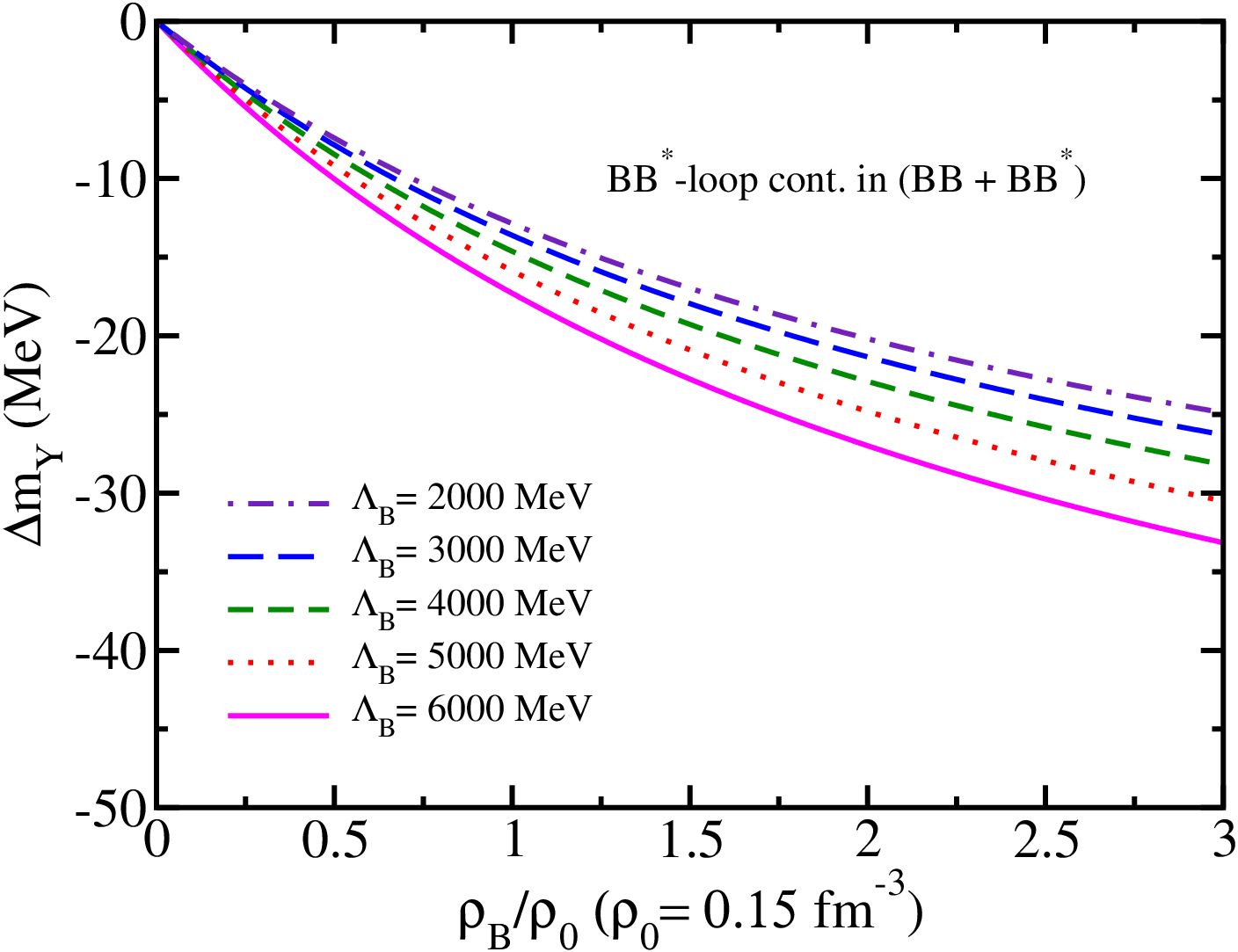}
\\
\vspace{6ex}
\includegraphics[width=6.5cm]{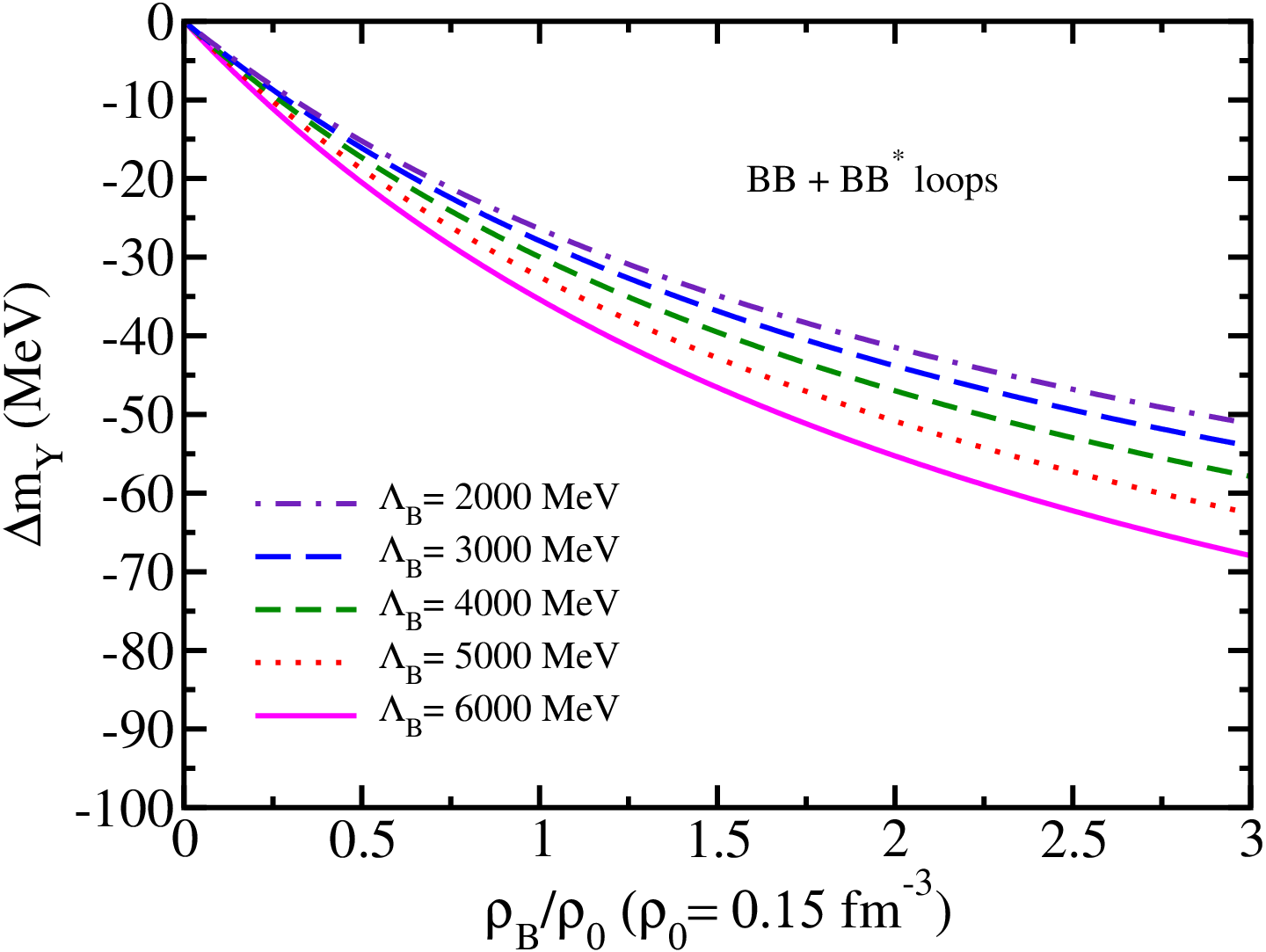}
\caption{$BB$ (top left), $BB^{*}$ (top right) and total (bottom) meson loop contributions to the 
$\Upsilon$ mass shift 
versus nuclear matter density for five different values of the cutoff 
mass $\Lambda_{B} (=\Lambda_{B^*})$.}
\label{fig4}
\end{figure}

\begin{figure}[htb]%
\vspace{2ex}
\centering
\includegraphics[width=6.5cm]{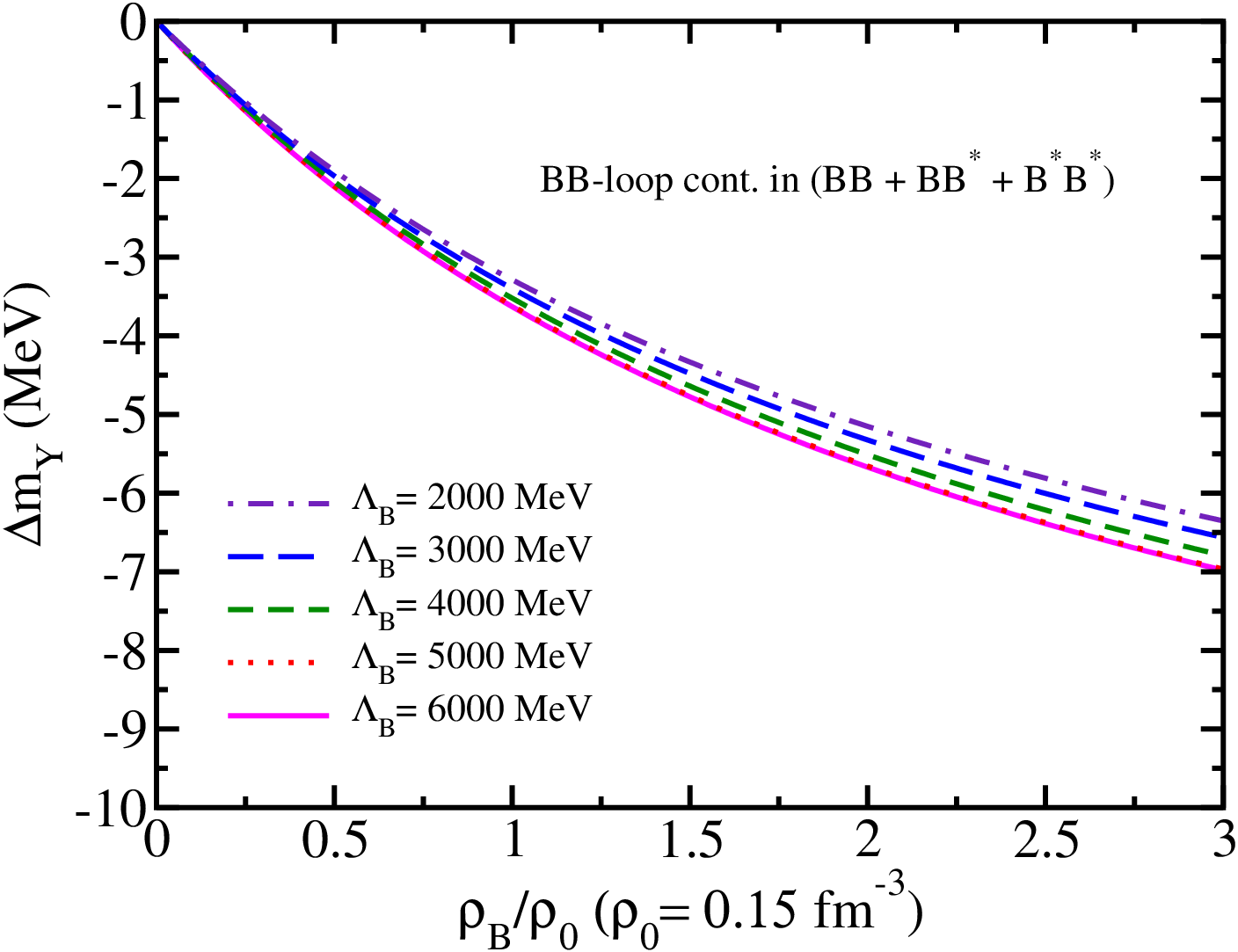}
\hspace{2ex}
\includegraphics[width=6.5cm]{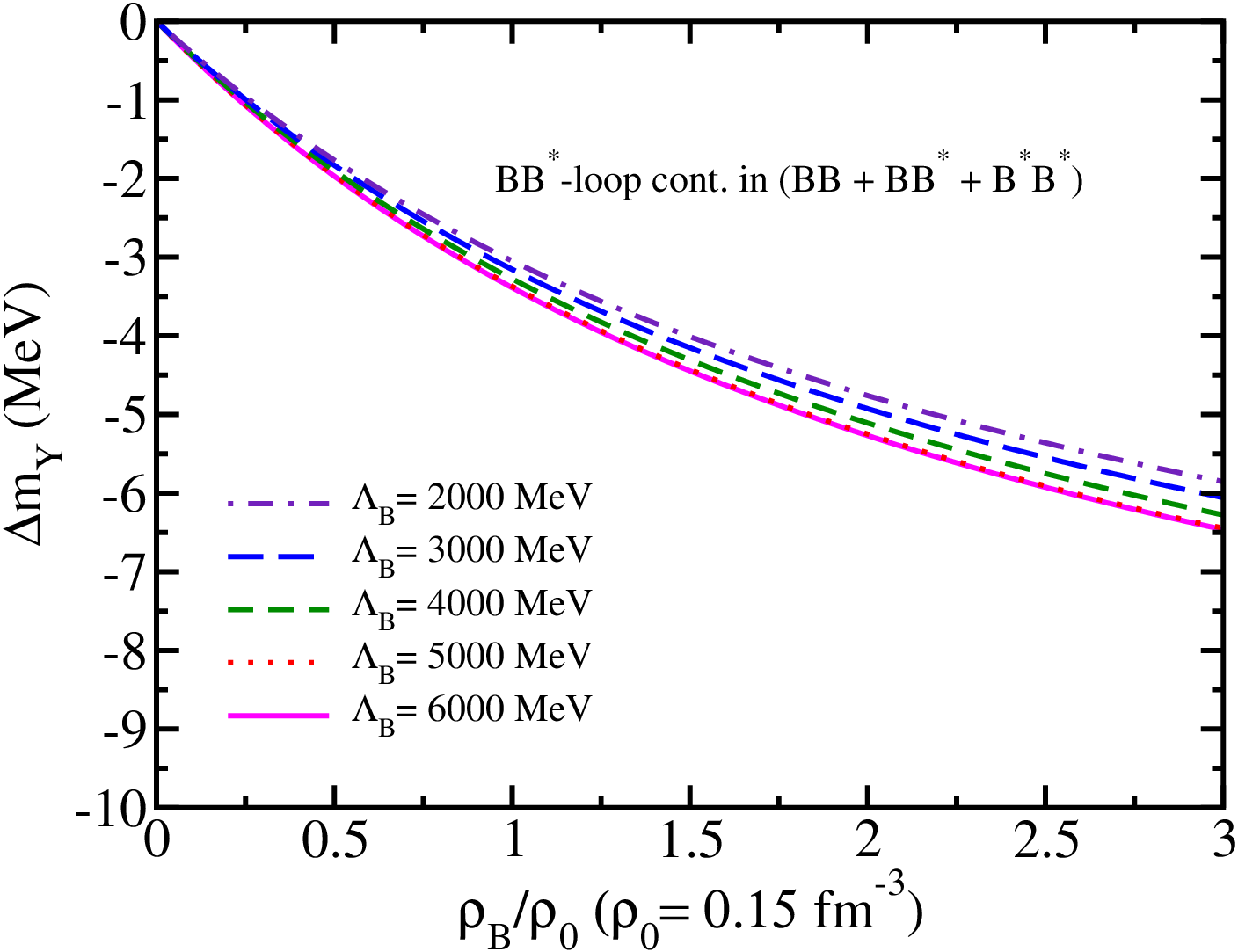}
\\
\vspace{6ex}
\includegraphics[width=6.5cm]{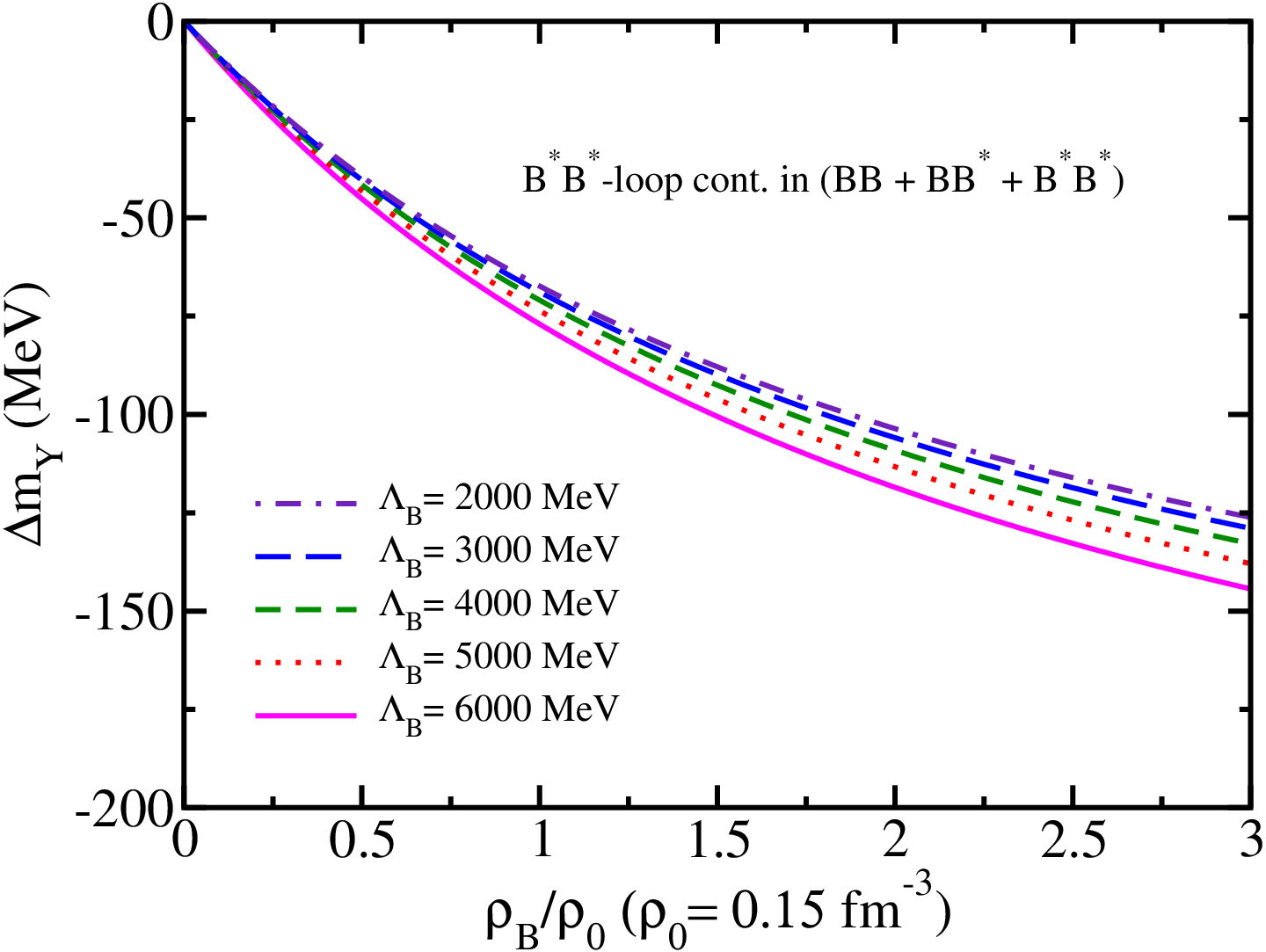}
\hspace{2ex}
\includegraphics[width=6.5cm]{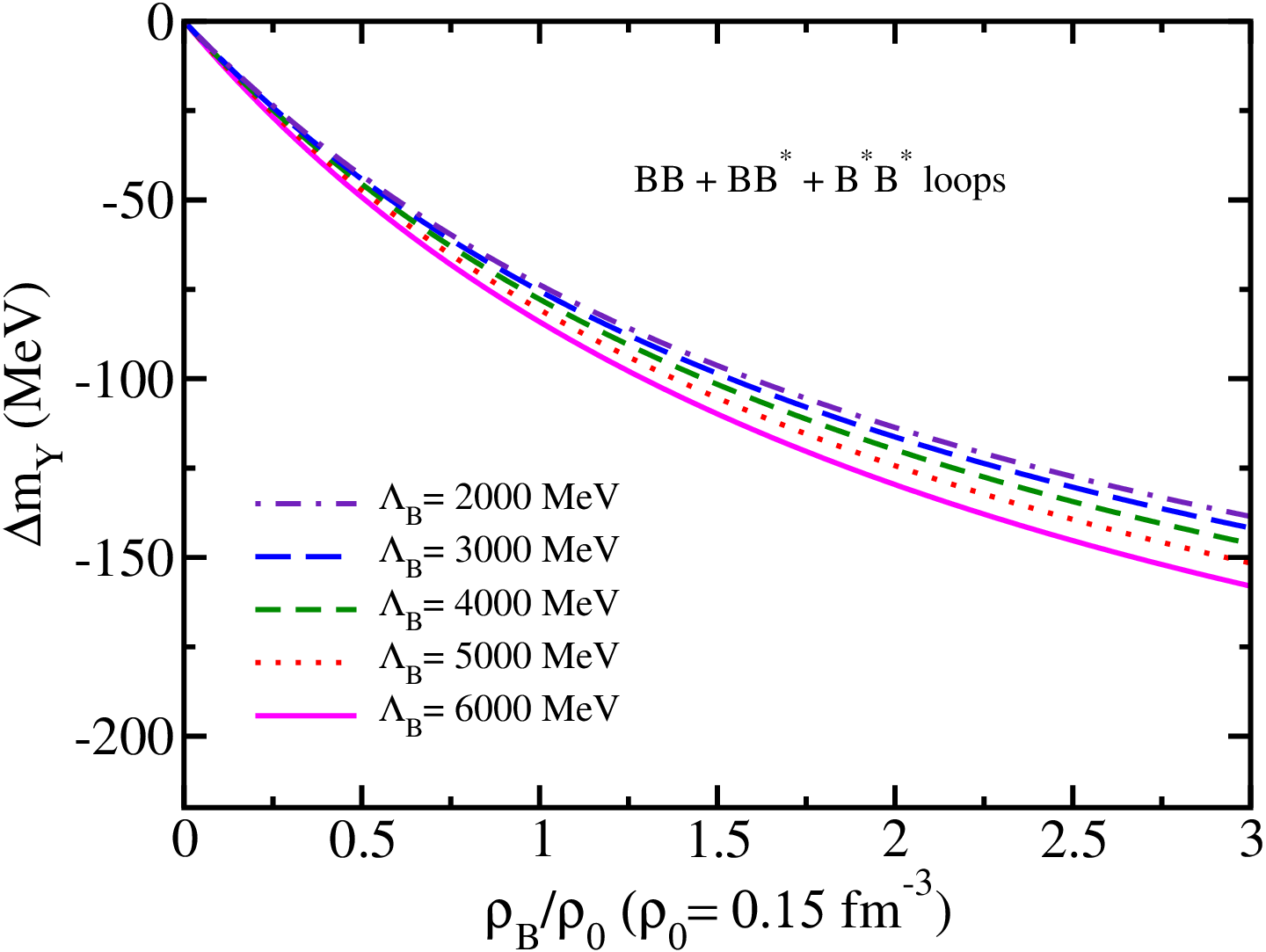}
\caption{$BB$ (top left), $BB^{*}$ (top right), $B^{*}B^{*}$ (bottom left) and
total (bottom right) meson loop contributions to the $\Upsilon$ mass shift 
versus nuclear matter density for five different values of the cutoff mass 
$\Lambda_{B} (=\Lambda_{B^*})$.}%
\label{fig5}%
\end{figure}

In Fig.~\ref{fig3} we show the $\Upsilon$ mass shift, taking the total contribution 
to be the $BB$ meson loop for five values of the cutoff mass $\Lambda_{B}$, 
2000, 3000, 4000, 5000 and 6000 MeV 
(these values will be applied for all the studies in the following with $\Lambda_B = 
\Lambda_{B^*}$). 
As one can see, the effect of the decrease in the $B$ meson in-medium mass    
yields a negative mass shift of the $\Upsilon$. 
The decrease of the $B$ meson mass in (symmetric) nuclear matter enhances the $BB$ meson loop 
contribution, thus the self-energy contribution in the medium becomes larger 
than that in the free space. 
This negative shift of the $\Upsilon$ mass is also dependent on the
value of the cutoff mass $\Lambda_B$, i.e., the amount of the mass shift 
increases as $\Lambda_B$ value increases, ranging 
from -16 to -22 MeV at the symmetric nuclear matter 
saturation density, $\rho_0 = 0.15$ fm$^{-3}$.

Next, in Fig.~\ref{fig4} we show the $\Upsilon$ mass shift taking the total self-energy 
contribution to be the $(BB + BB^*)$ meson loops. 
The contributions are shown for the $BB$ meson loop (top left), 
$BB^{*}$ meson loop (top right), and the total $(BB + BB^*)$ meson loops (bottom). 
The total mass shift at $\rho_0$ ranges from -26 to -35 MeV for 
the same range of the $\Lambda_B (= \Lambda_{B^*})$ values. 

Finally, we show in Fig.~\ref{fig5} the $\Upsilon$ mass shift 
taking the total self-energy contribution to be the $(BB + BB^* + B^*B^*)$ meson loops. 
The contributions are shown for the $BB$ meson loop (top left), 
the $BB^{*}$ meson loop (top right), $B^*B^*$ meson loop (bottom left),  
and the total $(BB + BB^* + B^*B^*)$ meson loops (bottom right).
The total mass shift at $\rho_0$ ranges from -74 to -84 MeV 
for the same range of the $\Lambda_B (= \Lambda_{B^*})$ values.

It is important to note that due to the unexpectedly    
larger contribution from the heavier meson-pair $B^{*}\overline{B^{*}}$   
meson loop ($B^*B^*$ meson loop) to the $\Upsilon$ mass shift than the other 
lighter-meson-pair loops $BB$ and $BB^*$   
presented in Fig.~\ref{fig5}, we regard the form factor used for 
the vertices in the $B^{*}B^{*}$ meson loop may not be appropriate, 
and need to consider either different 
form factors, or adopt an alternative regularization 
method in the future.
\vspace{1ex}

\noindent
{\it Summary for the $\Upsilon$ mass shift:}\\ 
The $\Upsilon$ mass shift is shown separately in Figs.~\ref{fig3},~\ref{fig4} 
and ~\ref{fig5}, by the difference in the intermediate states contributing for 
the total $\Upsilon$ self-energy, namely, by the $BB$, 
$\left(BB + BB^{*}\right)$, and $\left(BB + BB^{*} + B^{*}B^{*}\right)$ meson 
loops. The corresponding $\Upsilon$ mass shift at $\rho_0$ ranges, 
(-16 to -22) MeV, (-26 to -35) MeV, and (-74 to -84) MeV, for the adopted range 
of the $\Lambda_B (=\Lambda_{B^*}$) values.
The results indicate that the dependence on the values of the cutoff 
mass $\Lambda_B (=\Lambda_{B^*})$ is rather small compared to that of the $\Lambda_D 
(=\Lambda_{D^*})$ for the $J/\Psi$ case as will be discussed later, 
and this gives smaller ambiguities for our prediction 
originating from the cutoff mass values.

\subsection{Comparison with $J/\Psi$ mass shift}
\label{comp}

\begin{figure}[htb]
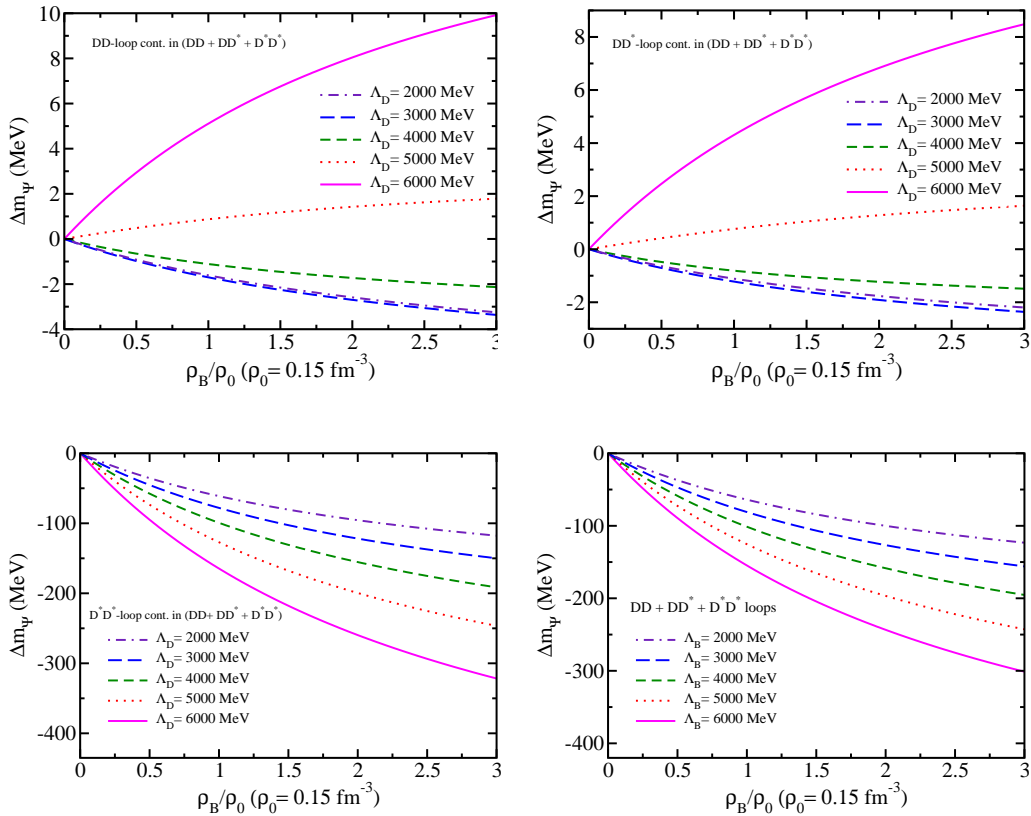
%
\vspace{2ex}
\centering
\includegraphics[width=6.5cm]{3_DD.eps}
\hspace{2ex}
\includegraphics[width=6.5cm]{3_DDs.eps}
\\
\vspace{6ex}
\includegraphics[width=6.5cm]{3_DsDs.eps}
\hspace{2ex}
\includegraphics[width=6.5cm]{3_Jtotal.eps}
\caption{$DD$ (top left), $DD^{*}$ (top right), $D^{*}D^{*}$ (bottom left) and
total (bottom right) meson loop contributions to the $J/\Psi$ mass shift 
versus nuclear matter density for five different values of the cutoff mass 
$\Lambda_D (=\Lambda_{D^*})$.}%
\label{fig6}%
\end{figure}

The issue of the larger contribution from the heavier vector meson loop,  
in the present case $B^{*}B^{*}$ meson loop, was already observed 
in a previous study of the $J/\Psi$ 
mass shift due to the heavier meson $D^{*}D^{*}$ loop contribution, where a 
similar nongauged effective Lagrangian was used and 
no cutoff readjustment was made for the heavier vector meson intermediate 
states~\cite{Krein:2010vp}. 
The cutoff mass value readjustment in a proper manner is important, 
because it controls the fluctuations from the shorter distances.
However, we do not try this in the present exploratory study, 
since we first need to see the bare result without 
readjusting, so that we are able to compare with those of the $J/\Psi$ case, 
focusing on the heavy quark and heavy meson symmetry.

We have calculated the total ($DD + DD^* + D^*D^*$) meson loop contribution  
for the $J/\Psi$ mass shift as featured in Ref.~\cite{Krein:2017usp}  
using the same effective Lagrangian and cutoff mass values to compare with the  
total ($BB + BB^* + B^*B^*$) meson loop contribution in the $\Upsilon$ mass shift.
The free space masses of the $J/\Psi$, $D$ and $D^{*}$ mesons used are
3097, 1867 and 2009 MeV~\cite{PDG2020}, respectively.  
The result is presented in Fig.~\ref{fig6}.  
The $D^{*}D^{*}$ meson loop contribution 
for the $J/\Psi$ mass shift ranges from -61 to -164 MeV at $\rho_0$, 
which is mostly larger than that of the $B^{*}B^{*}$ (-67 to -77 MeV at $\rho_0$) 
for the same range of the cutoff mass values in the corresponding form factors. 
Note that, the larger cutoff mass values, $\Lambda_D (=\Lambda_{D^*}) = 5000$ and $6000$ MeV,  
may not be appropriate as will be discussed in the following.
We can see from Fig.~\ref{fig6} that the  
closer the cutoff mass value gets to the $J/\Psi$ mass, 
less pronounced the negative mass shift becomes, 
until it reaches a transition point (when $\Lambda_{D}$ is larger than 
the $J/\Psi$ free space mass), where the potential starts to become even positive. 
Naively, according to the second order perturbation theory in quantum mechanics, 
they should give the negative contribution, but the positive contributions 
for $\Lambda_D = 5000$ and $6000$ MeV, thus suggest that such larger values of 
the cutoff mass may not be justified for the form factor used.
One can expect a similar behavior in the $BB$ meson loop in the 
total $(BB+BB^*+B^*B^*)$ meson loop contribution when the cutoff mass value gets 
closer to the $\Upsilon$ mass.
Indeed, such behavior is observed for the $BB$ and $BB^*$ meson loop contributions, 
for the cutoff mass values larger than 10000 MeV.
As already commented in Subsec.~\ref{uslef}, the large cutoff-mass values than 
the corresponding vector meson mass means that the distance for the interaction between 
the vector meson and the intermediate state meson included is shorter than the meson 
overlapping region, and a physical picture as an isolated vector meson 
is lost --- one also needs to consider the quark-quark, quark-antiquark, and antiquark-antiquark  
interactions and/or the corresponding correlations at the quark level  
in such short distances, where the present approach does not have. 

The bad high-energy behavior of the vector meson propagator is well known. 
To evaluate amplitudes in high-energy region that contain vector meson propagators 
in spontaneously broken gauge theory such as the weak interaction in the Standard Model, 
the $R_{\xi}$ gauge is usually used. The $R_{\xi}$ gauge with 
$\xi = 1$ ('t Hooft-Feynman gauge) makes the high-energy behavior of the vector meson propagators 
similar to that of the spin-0 meson 
propagators~\cite{tHooft:1971qjg,tHooft:1971akt,Lee:1971kj,Fujikawa:1972fe}. 
$R_{\xi}$ gauge removes unphysical degrees of freedom associated 
with the Goldstone bosons. In the present case, we cannot justify to use such vector meson 
propagators, so we need to tame the bad high-energy behavior phenomenologically. 
We can do this by introducing a phenomenological form factor for the $BB$ meson loop case.  
But for the $BB^{*}$ and $B^{*}B^{*}$ meson loops we simply discard their contributions 
in the present study as was practiced in Ref.~\cite{Krein:2017usp}. 
Therefore, our prediction should be regarded based on the minimum contribution  
with respect to the intermediate state meson loops, namely by only the 
$BB$ meson loop contribution as in Ref.~\cite{Krein:2017usp}, 
which took only the $DD$ meson loop contribution for estimating the $J/\Psi$ mass shift.
Regarding the form factors, another choice of form factors is possible 
to moderate the high-energy behavior~\cite{Gryniuk:2020mlh,Tsushima:1994rj,Lin:1999ad}, 
and an initial study of using a different form factor will be performed in Sec.~\ref{diff_ff}.

Furthermore, although we have chosen the same coupling constants for 
$\Upsilon BB$, $\Upsilon BB^{*}$, and 
$\Upsilon B^{*}B^{*}$, it is certainly possible to use the  
different values for the coupling constants. 
Some studies of SU(4) flavor symmetry breaking couplings in charm sector offer 
alternative ways for the calculation of these coupling constants. 
This can be extended to include SU(5) symmetry 
breaking couplings. But for the flavor SU(5) sector, the breaking 
effect is expected to be even larger than that for the SU(4) sector, since bottom quark mass is 
much heavier than the charm quark, and the SU(5) symmetry breaking 
is expected to be larger. 
There are some studies focused on the SU(4) symmetry 
breaking of the coupling 
constants, although the results are not conclusive.  
A recent calculation~\cite{Lucha:2015dda} used dispersion formulation of the relativistic 
constituent quark model, where the couplings were 
obtained as residues at the poles of suitable form factors. 
Two other studies are made by the Schwinger-Dyson-equation-based  
approaches for QCD~\cite{ElBennich:2011py,El-Bennich:2016bno}. 
In the both approaches, the obtained results for the SU(4) symmetry 
breaking are considerably larger than those obtained using QCD sum-rule approach. 
We plan to do more dedicated studies on the issues in the future.
In the present study, the coupling constant $g_{\Upsilon BB}=13.2$ contains 
SU(5) symmetry breaking effect with respect to that of the corresponding 
charm sector, $g_{J/\Psi DD}=7.64$, where both of them are determined 
using the VMD model with experimental data.

We emphasize again that, the prediction for the $J/\Psi$ mass shift   
made solely by the $DD$ meson loop, gives -3.0 to -6.5 MeV based on 
Refs.~\cite{Krein:2010vp,Krein:2017usp} (-5 to -21 MeV for the same range 
of the cutoff $\Lambda_D$ value, 2000 to 6000 MeV), while for the $\Upsilon$ mass shift, 
taking only the contribution from the $BB$ meson loop, gives -16 to -22 MeV.
In Sec.~\ref{hlimit} we will make some study for the $\Upsilon$ and $\eta_b$ 
mass shifts focusing on the SU(5) symmetric coupling constant   
between the charm and bottom sectors, as well as a coupling constant 
in a broken SU(5) symmetry scheme between the $\Upsilon$ and $\eta_b$.

One might question further, as to why the $\Upsilon$ ($\eta_b$) mass shift is larger than 
that of the $J/\Psi$ ($\eta_c$), although we have already commented 
the main reason by the larger coupling constant obtained by 
the VMD model with the experimental data. 
(The other way, why the bottom sector coupling constant is larger than that of the charm sector, 
or the corresponding experimental data in free space to determine the coupling constant is larger.)
Of course, the heavier $B$ and $B^*$ meson masses 
than the corresponding $D$ and $D^*$ meson masses also 
influence the $\Upsilon$ and $J/\Psi$ mass shift difference, 
although the heavier $B$ and $B^*$ meson masses counteract to reduce 
the $\Upsilon$ mass shift, since the heavier particles are more difficult 
to be excited in the intermediate states of the $\Upsilon$ self-energy meson loops.
To understand better, let us consider the systems of the bottom and charm sectors, 
$(D,D^*)$ and $(B,B^*)$ meson systems. 
For these two sets of systems, 
we can estimate the difference in the (heavy quark)-(light quark) interactions 
by $(m_D-m_c, m_{D^*}-m_c)$ and $(m_B-m_b, m_{B^*}-m_b)$, 
since the existence of the light quark and its interaction with the heavy quark 
in each system gives the total mass of each meson. 
Using the values (all in MeV in the following), $m_c=1270, m_D=1870, m_{D^*}=2010$, 
$m_b=4180, m_B=5279$, and $m_{B^*}=5325$, 
we get $(m_D-m_c, m_{D^*}-mc)=(600,740)$ 
and $(m_B-m_b, m_{B^*}-m_b)=(1099,1145)$.
These results indicate that the $b$-(light quark) interaction 
is more attractive than that of the $c$-(light quark),
since the larger mass differences for the $b$-quark sector mesons without light quarks 
than those corresponding for the $c$-quark sector mesons, are diminished more than those for the 
corresponding $c$-quark sector mesons as the experimentally observed masses --- the consequence of 
more attractive $b$-(light quark) interaction.
This implies that the bottomonium-nucleon (bottomonium-(nuclear matter)) interaction 
is more attractive than that of the charmonium-nucleon 
(charmonium-(nuclear matter)). 
In this way, we may be able to understand the larger mass shift of  
the $\Upsilon$ ($\eta_b$) than that of the $J/\Psi$ ($\eta_c$)  
due to the interaction with the nuclear medium --- composed 
of infinite number of light quarks.

\section{{\large\boldmath{$\eta_{\MakeLowercase{b}}$}} mass shift}
\label{emass}

Based on the discussion and analysis made for the $\Upsilon$ mass shift 
so far, we proceed to study the $\eta_b$ mass shift.
By the same philosophy as adopted for the $\Upsilon$ mass shift, 
we take only the $BB^*$ meson loop contribution for the $\eta_b$ self-energy 
as our prediction, 
namely, participants in the self-energy diagram of the $\eta_b$ meson are, one 
vector meson $B^*$, and two pseudoscalar mesons $\eta_b$ and $B$.

Before going into the details of the $\eta_b$ mass shit, 
we comment on an issue discussed in the pionic-atom study, 
the Ericson-Ericson-Lorentz-Lorenz (EELL) double (multiple) scattering 
correction~\cite{Ericson:1966fm,Ericson,Brown:1990wyp}. 
Since the mean field potentials become constant and the coupling constants 
are determined within the Hartree approximation (local) in the present QMC model treatment,   
the EELL double scattering correction (nonlocal effect), which was also considered 
for the $\eta$ and $\eta'$ meson mass shifts in nuclear matter~\cite{Bass:2005hn}, 
may be regarded as effectively included in our calculation.
In fact, based on this argument with some discussions, the EELL effect was not 
included explicitly in the study of the $\eta_c$ mass shift~\cite{Cobos-Martinez:2020ynh}. 
We simply follow Ref.~\cite{Cobos-Martinez:2020ynh} on the issue of the EELL effect 
in the present study.
Aside from this, we mention that there is a lack of useful information in the literature 
on the $\eta_b$-nucleon scattering length, even if one wants to estimate  
the EELL effect. 

The effective Lagrangian for the $\eta_b BB^*$ interaction is obtained from Eq.~(\ref{Lint}) 
in the same way as those for the $\Upsilon$, and we get,  
\begin{eqnarray}
{\cal L}_{\eta_b BB^*} 
&=& i g_{\eta_b BB^*}
\left\{ (\partial^\mu \eta_b) 
\left( \overline{B^*}_\mu B - \overline{B} B^*_\mu \right)
- \eta_b 
\left[ \overline{B^*}_\mu (\partial^\mu B) - (\partial^\mu \overline{B}) B^*_\mu \right]
\right\}, 
\label{Letab1}
\end{eqnarray}
where the coupling constant in the SU(5) scheme is used 
for $g_{\eta_b BB^*}$: 
\begin{equation}
g_{\eta_b BB^*} = g_{\Upsilon BB} = g_{\Upsilon B^*B^*} = \frac{5g}{4\sqrt{10}}.
\end{equation}

We also study the anomalous coupling $\eta_bB^*B^*$ contribution,
\begin{equation}
{\cal L}_{\eta_b B^*B^*} 
= \frac{g_{\eta_b B^*B^*}}{m_{\eta_b}}\varepsilon_{\alpha \beta \mu \nu}
\eta_b \left[ (\partial^{\alpha} \overline{B^*}^\beta) (\partial^{\mu}B^{* \nu}) \right], 
\label{Letab2} 
\end{equation}
assuming 
$g_{\eta_b B^*B^*} \left(\frac{m_{\eta_b}}{m_\Upsilon}\right) = g_{\eta_b BB^*}
(= g_{\Upsilon B^{*}B^{*}})$.
If we rely on the heavy quark symmetry and/or heavy meson (spin) symmetry, 
the above relation, $g_{\eta_b BB^*} = g_{\Upsilon BB}$, which is used for our  
prediction of the $\eta_b$ mass shift, may be justified.

The $\eta_b$ self-energy is expressed by~\cite{Cobos-Martinez:2020ynh} 
\bea
\hspace{-40ex}\Sigma_{\eta_b} = \Sigma_{\eta_b}^{BB^*} &+& \Sigma_{\etab}^{B^*B^*},   
\eea
with
\bea
\Sigma_{\eta_b}^{BB^*} 
&=& \frac{8 g_{\etab B B^*}^{2}}{\pi^{2}}\int_{0}^{\infty}
    \dx{q} \tbf{q}^{2} \tilde{F}_{BB^*}(\tbf{q}^2) I_{BB^*}(\tbf{q}^{2}),  
\label{SigetabBBs}
\\
\Sigma_{\etab}^{B^*B^*}(\tbf{q}^{2})
&=& \frac{2g_{\etab B^* B^{*}}^{2}}{\pi^{2}}\int_{0}^{\infty}
    \dx{q} \tbf{q}^{4} \tilde{F}_{B^*B^*}(\tbf{q}^2) I_{B^*B^*}(\tbf{q}^{2}),  
\label{SigetabBsBs}
\eea
and for the $\eta_b$ at rest, 
\bea
I_{BB^*}(\tbf{q}^{2})
&=& \left. \frac{m_{\etab}^{2}(-1+(q^0)^2/m_{B^{*}}^{2})}
{(q^{0}+\omega_{B^{*}})(q^{0}-\omega_{B^*}) 
(q^{0}-m_{\etab}-\omega_{B})}\right|_{q^{0}=m_{\etab}- \omega_B} 
\nonumber \\
&&\hspace{5ex} + \left. \frac{m_{\etab}^{2}(-1+(q^0)^2/m_{B^{*}}^{2})}
{(q^{0}-\omega_{B^{*}})(q^{0}-m_{\etab}+\omega_{B}) 
(q^{0}-m_{\etab}-\omega_{B})}\right|_{q^{0}=-\omega_{B^*}},  
\label{IBBs}
\\
\\
I_{B^*B^*}(\tbf{q}^{2})
&=& \left. \frac{1}{(q^0 + \omega_{B^*})(q^0 -\omega_{B^*}) 
(q^{0}-m_{\etab} -\omega_{B^*})} \right|_{q^0=m_{\etab}-\omega_{B^*}}  
\nonumber \\
&&\hspace{5ex} + \left. \frac{1}{(q^0-\omega_{B^*})(q^0-m_{\etab} + \omega_{B^*}) 
(q^0-m_{\etab} - \omega_{B^*})} \right|_{q^0=-\omega_{B^*}},
\label{IBsBs}
\eea
and $\omega_{B,B^*}=(\tbf{q}^{2}+m_{B,B^*}^{2})^{1/2}$.
For the $\etab BB^*$ and $\etab B^*B^*$ vertices, we use the similar form factors
as for the $\Upsilon$ case, 
$\tilde{F}_{BB^*}(\tbf{q}^2) = \tilde{u}_B(\tbf{q}^2) \tilde{u}_{B^*}(\tbf{q}^2)$ 
and $\tilde{F}_{B^*B^*}(\tbf{q}^2) = \tilde{u}_{B^*}^2(\tbf{q}^2)$, 
respectively, with
\begin{equation}
\label{eqn:FF}
\tilde{u}_{B,B^*}(\tbf{q}^{2})=
\left(
\frac{\Lambda_{B,B^*}^{2} + m_{\etab}^{2}}{\Lambda_{B,B^*}^{2}
    +4\omega_{B,B^*}^{2}(\tbf{q}^{2})}
\right)^{2}.
\end{equation}
Note that, in Eq.~(\ref{IBsBs}) no terms arise  
originating from $\propto (q^\lambda q^\sigma/m_{B^*}^2)$ in the $B^*$ propagators, 
due to the (two multiplication of) totally antisymmetric $\varepsilon$ tensor in the amplitude.
Thus, in contrast to the $B^*B^*$ loop contribution for the $\Upsilon$ self-energy, 
the $B^*B^*$ loop contribution for the $\eta_b$ self-energy is expected to be small, 
and to give a less divergent high energy behavior than 
that in the $\Upsilon$ self-energy.

\subsection{Results for {\large$\eta_b$} mass shift}
\label{eres}

\begin{figure}[htb]
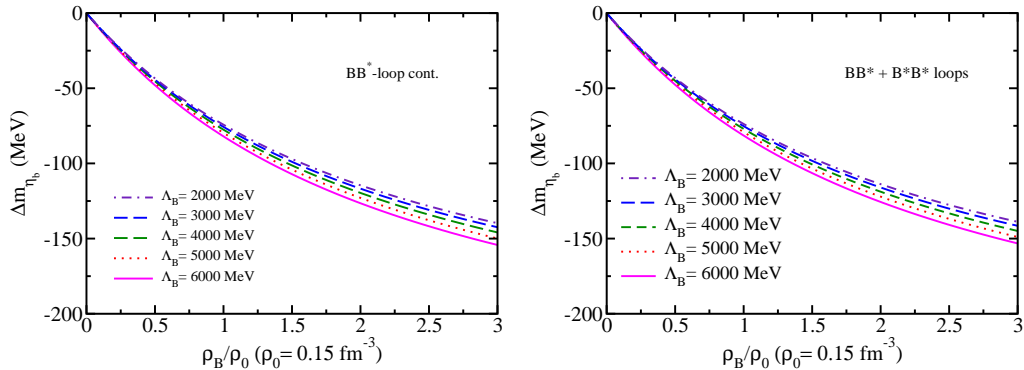
%
\vspace{4ex}
\centering
\includegraphics[width=6.5cm]{etab_BBs.eps}%
\hspace{2ex}
\includegraphics[width=6.5cm]{etab_total.eps}
\caption{$\eta_b$ mass shift from the $BB^*$ 
(left panel) and $BB^*+B^*B^*$ (right panel) meson loops  
versus nuclear matter density for five different values of the cutoff mass $\Lambda_{B} 
(=\Lambda_{B^*})$.}%
\label{fig7}%
\end{figure}

To be consistent, we show in Fig.~\ref{fig7} the calculated $\eta_b$ mass shift for including only 
the $BB^*$ loop --- our prediction (left panel), and that including the 
($BB^*+B^*B^*$) loops (right panel) for five different cutoff mass values, 
the same as those applied for the $\Upsilon$.
The $\eta_b$ mass shift at $\rho_0$ with the $BB^*$ loop only (left panel) 
ranges from -75 to -82 MeV, while that with the ($BB^*+B^*B^*$) loops (right panel) 
ranges from -74 to -81 MeV , where we have used $m_{\eta_b} = 9399$ MeV~\cite{PDG2020} 
for the free space value. The two results show very similar mass shifts.
In the latter case, the $B^*B^*$ loop contribution ranges from +3 to +5 MeV at $\rho_0$.
The reason for the smaller contribution compared to that for the $\Upsilon$ self-energy 
is explained already in Subsec.~\ref{emass}.
By the above difference and the fact that the smallness of the $B^*B^*$ loop 
contribution for the $\eta_b$ self-energy, we can conclude that the large $B^*B^*$ loop 
contribution for the $\Upsilon$ self energy arises due to (summation of) the $\Upsilon$ 
polarization vector, correlating with the momentum dependent part in numerators of the two $B^*$ 
propagators, $\propto (q^\mu q^\nu / m_{B^*}^2)$. The part also often gives a divergent high energy 
behavior with including the integral measure $\int d^4 q$ in the $B^*$ meson propagator.

Note that, similar to the $\Upsilon$ mass shift, dependence of the $\eta_b$ mass shift 
on the cutoff mass value $\Lambda_B (=\Lambda_{B^*})$ is again small, 
and it gives less ambiguity for the prediction originating from the cutoff mass value.
Unexpectedly, the $\eta_b$ mass shift is much larger than the  
predicted $\Upsilon$ mass shift due to only the $BB$ meson loop contribution, 
although the same lowest order $BB^*$ meson loop contribution 
(one vector and two pseudoscalar mesons) is included in the self-energy with 
the same range of the cutoff mass $\Lambda_B 
(=\Lambda_{B^*})$ values. 
One of the main reasons lies in the Lagrangian  
Eq.~(\ref{Letab1}). By the explicit calculation one can show that  
the large number of the interaction terms in the Lagrangian   
contributes to the self-energy, results to make 
the total contribution large. This is reflected in the coefficient in 
Eq.~(\ref{SigetabBBs}), and in contrast to the case of the $BB$ meson loop contribution 
in the $\Upsilon$ self-energy.
The similar, larger mass shift of the $\eta_c$ than that of the $J/\Psi$ 
was also observed in Ref.~\cite{Cobos-Martinez:2020ynh}, 
using the corresponding Lagrangians in the SU(4) sector.

\section{Heavy quark (heavy meson) symmetry limit}
\label{hlimit}

In the following, we consider a heavy quark (heavy meson) symmetry 
limit, by treating the $\Upsilon$ and $J/\Psi$, as well as the $\eta_b$ and 
$\eta_c$ mesons on the same footing, namely, to assign the 
same coupling constant value in the corresponding interaction vertices 
with $g_{J/\Psi DD}=g_{\eta_c DD^*}=7.64$  
used in Refs.~\cite{Krein:2010vp,Cobos-Martinez:2020ynh}. 
Furthermore, to compare with the $\eta_c$ mass shift given in Ref.~\cite{Cobos-Martinez:2020ynh} 
calculated by considering an SU(4) symmetry breaking by   
$g_{\eta_c DD^*}=(0.6/\sqrt{2})\,g_{J/\Psi DD} \simeq 0.424\,g_{J/\Psi DD}$, 
we also study the same case for the $\eta_b$ mass shift.

\begin{figure}[htb]
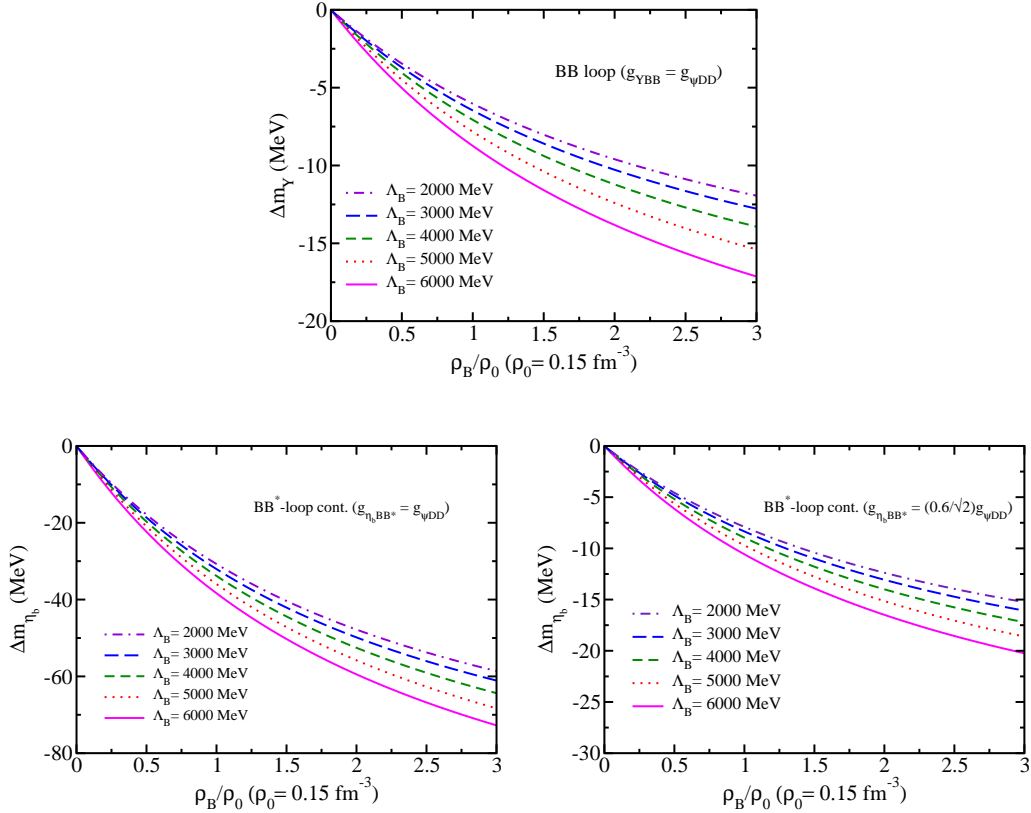
%
\centering
\includegraphics[width=6.5cm]{BB_gJpsi.eps}
\vspace{6.2ex}
\\
\includegraphics[width=6.5cm]{BBs_gJpsi.eps}
\hspace{2ex}
\includegraphics[width=6.5cm]{BBs_gJpsi_06sqr.eps}
\caption{$\Upsilon$ (top) and $\eta_b$ (bottom left) mass shifts 
calculated using the coupling constant relation 
$g_{\Upsilon BB}=g_{\eta_b BB^*}=g_{\eta_c DD^*}=g_{J/\Psi DD}$, and 
that of the $\eta_b$ calculated by a broken SU(5) symmetry, 
$g_{\eta_b BB^*}=(0.6/\sqrt{2})\,g_{\eta_c DD^*}=(0.6/\sqrt{2})\,g_{J/\Psi DD}$ (bottom right), 
with $g_{J/\Psi DD}=7.64$~\cite{Krein:2010vp,Cobos-Martinez:2020ynh}.
}
\label{figHQsym}
\end{figure}

In Fig.~\ref{figHQsym} we show the mass shifts calculated in the heavy quark (heavy meson) 
symmetry limit, and also the broken SU(5) symmetry in this limit 
for the $\eta_b$, namely,  
(i) $\Upsilon$ mass shift calculated by the coupling constant appearing in the 
self-energy by $g_{\Upsilon BB}=g_{J/\Psi DD}$ (top), 
(ii) $\eta_b$ mass shift by $g_{\eta_b BB^*}=g_{\eta_c DD^*}=g_{J/\Psi DD}$ (bottom left), 
and (iii) $\eta_b$ mass shift with a broken SU(5) symmetry by  
$g_{\eta_b BB^*}=(0.6/\sqrt{2})\,g_{\eta_c DD^*}=(0.6/\sqrt{2})\,g_{J/\Psi DD}$ (bottom right), 
where we use $g_{J/\Psi DD}=7.64$~\cite{Krein:2010vp,Cobos-Martinez:2020ynh}.

Although the coupling constant value used for the bottom sector is now the same as that for 
the charm sector, since the relevant mesons in the bottom sector are heavier than 
those of the corresponding charm sector in free space as well as in medium, the 
dependence of the mass shifts on the cutoff mass value becomes more insensitive than 
that of the charm sector for the same range of 
the cutoff mass values. Thus, the ranges of the mass shifts of $\Upsilon$ and $\eta_b$ 
by the values of the cutoff mass becomes smaller than those of the corresponding 
$J/\Psi$ and $\eta_c$. 
In this limit, the obtained mass shifts range at $\rho_0$ corresponding to the cases 
stated above are,  
(i) -6 to -9 MeV (-5 to -21 MeV) for $\Upsilon$ ($J/\Psi$),  
(ii) -31 to -38 MeV (-49 to -87 MeV) for $\eta_b$ ($\eta_c$), 
and (iii) -8 to -11 MeV (-17 to -51 MeV) for $\eta_b$ ($\eta_c$), 
with the same range of the cutoff mass values $\Lambda_B$ from 2000 to 6000 MeV 
($\Lambda_D$ from 2000 to 6000 MeV).
These results indicate, as one can expect, the amounts of mass shifts for the 
$\Upsilon$ and $\eta_b$ become smaller than those of the corresponding $J/\Psi$ and $\eta_c$.
This fact confirms that the larger mass shifts of the $\Upsilon$ and $\eta_b$ 
than those of the $J/\Psi$ and $\eta_c$ obtained in previous sections are 
due to the larger coupling constant $g_{\Upsilon BB} = 13.2$ than   
$g_{J/\Psi DD} = 7.64$, where both values are obtained by the VMD model 
with the corresponding experimental data.
If this heavy quark (heavy meson) symmetry limit is more closely realized in nature, 
we expect to obtain smaller mass shifts for the $\Upsilon$ and $\eta_b$ than those of the 
corresponding $J/\Psi$ and $\eta_c$.

\section{Initial study of using a different form factor}
\label{diff_ff}

To see the effects of the form factor on the $\Upsilon$ and $\eta_b$ mass shifts, 
we calculate their mass shifts using a different form factor for  
the lowest order contributions, $BB$ and $BB^*$ loops, respectively (our predictions), 
as an initial study.
(We plan to perform an elaborate study for the effects of different form factors  
on the $\Upsilon$ and $\eta_b$ mass shifts.)
For this purpose we use the form factor~\cite{Tsushima:1991fe,Lin:1999ad,Lin:2000ke}, 
\begin{equation}
u_{B,B^{*}}(\textbf{q}^{2}) = \left(\frac{\Lambda^{2}_{\,\,B,B^{*}}}
{\Lambda^{2}_{\,\,B,B^{*}} + \textbf{q}^2}\right)^{2}, 
\label{newff}
\end{equation}
where the above $u_B$ and $u_{B^*}$ are applied in the same way as those already applied
for the corresponding vertices with $\Lambda_B=\Lambda_{B^*}$. 
The Fourier transform of the function $\Lambda_{\,B,B^*}^2/(\Lambda_{\,B,B^*}^2+{\bf q}^2)$  
in the form factor Eq.~(\ref{newff}) gives the Yukawa-potential type function, 
$\propto \exp(-\Lambda_{B,B^*}\,r)/r$, where $r$ is the distance of $B$ or $B^*$ meson 
from the $\Upsilon$ or $\eta_b$ meson for the corresponding vertices.
Derivative of the integrand $\Lambda_{\,B,B^*}^2/(\Lambda_{\,B,B^*}^2+{\bf q}^2)$ 
in the Fourier transform with respect to $\Lambda_{B,B^*}$ gives the 
dipole form Eq.~(\ref{newff}) aside the irrelevant constant, 
and the function $\propto \exp(-\Lambda_{B,B^*}\,r)$ structure  
remains and keeps controlling the same interaction range.
As is known, for $r > 1/\Lambda_B$, $\exp(-\Lambda_{B,B^*}\,r)$ 
suppresses effectively the interactions between the $\Upsilon$-$B$ and $\eta_b$-$B(B^*)$,  
where the $\Lambda_{\,B,B^*}$ dependence is expected to be more sensitive than 
the form factors in Eqs.~(\ref{ffups}) and~(\ref{eqn:FF}). 
Since the masses of $B$ and $B^*$ mesons are respectively $m_B = 5279$ MeV $\simeq 
m_{B^*} = 5325$ MeV, 
we expect that the cutoff mass value $\Lambda_{B,B^*} \simeq 5300$ MeV may be a reasonable 
value for the form factor Eq.~(\ref{newff}). 
Thus, for this form factor, we take the cutoff mass central value $\Lambda_B=5300$ MeV, and 
calculate the $\Upsilon$ and $\eta_b$ mass shifts for the cutoff-mass value range, 
4900 MeV $\le \Lambda_B \le$ 5700 MeV.

\begin{figure}[htb]
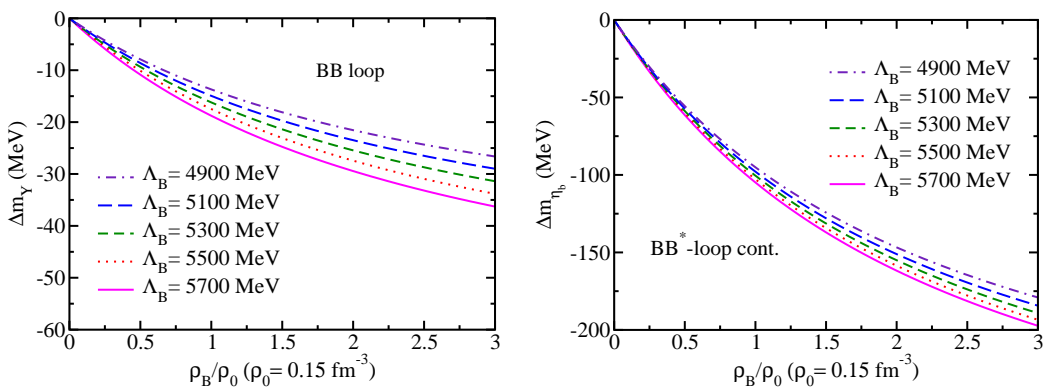
%
\vspace{1ex}
\centering
\includegraphics[width=6.5cm]{BB_dif_ff.eps}
\hspace{2ex}
\includegraphics[width=6.7cm]{BBs_dif_ff.eps}
\caption{$\Upsilon$ (left) and $\eta_b$ (right) mass shifts 
using a different form factor, 
$[\Lambda_{\,B,B^*}^2/(\Lambda_{\,B,B^*}^2+{\bf q}^2)]^2$ 
($\Lambda_B = \Lambda_{B^*}$),
with the cutoff-mass value range, 4900 MeV $\le \Lambda_B \le$ 5700 MeV.
}%
\label{figff}%
\end{figure}

The calculated mass shifts are shown in Fig.~\ref{figff}, 
for the $\Upsilon$ (left panel) and $\eta_b$ (right panel).
The results for the $\Upsilon$ and $\eta_b$ in Fig.~\ref{figff} may be 
compared with the corresponding results shown in Fig.~\ref{fig3} for 
the $\Upsilon$, and Fig.~\ref{fig7} the $\eta_b$, respectively.

The mass shifts at $\rho_0$ with the form factor Eq.~(\ref{newff}) are respectively, 
from -14 to -19 MeV for the $\Upsilon$, and from -95 to -104 MeV 
for the $\eta_b$ for the $\Lambda_B$ range, 4900 MeV $\le \Lambda_B \le$ 5700 MeV. 
The corresponding mass shifts at $\rho_0$ for the $\Upsilon$ with 
the form factor Eq.~(\ref{ffups}) and $\eta_b$ with the form factor Eq.~(\ref{eqn:FF}) are 
respectively, from -16 to -22 MeV, and -75 to -82 MeV for 
the cutoff mass range 2000 MeV $\le \Lambda_B \le$ 6000 MeV.
The different form factor given by Eq.~(\ref{newff}), which is more sensitive 
to the cutoff mass value, gives the similar mass shifts with those  
regarded as our predictions. 
The use of the form factors Eq.~(\ref{newff}) may give a better physical picture 
for the form factor.
The results shown in Fig.~\ref{figff}, especially for $\Upsilon$, 
have turned out to give very similar values 
with those corresponding ones obtained with the form factors  
Eqs.~(\ref{ffups}) but a wider range of the $\Lambda_B$, 
where the form factor Eq.~(\ref{newff}) can provide a better physical picture 
on the interaction range of the corresponding mesons.
As mentioned in Subsec.~\ref{ures}, 
we need, and plan to study further the effects of the other form factors, 
and/or the other regularization methods on the  
$\Upsilon$, $\eta_b$, $J/\Psi$, and $\eta_c$ mass shifts.

\section{Summary and Conclusion}
\label{con}

By extending the previous works, we have estimated for the first 
time the $B^*$, $\Upsilon$ and $\eta_b$ mass shifts 
in symmetric nuclear matter, neglecting any possible widths of the mesons.

For the $\Upsilon$, we have studied the $BB$, $BB^{*}$, and 
$B^{*}B^{*}$ meson loop contributions 
using effective SU(5) symmetry-based Lagrangians and the anomalous coupling one, 
with coupling constants calculated from the experimental data using 
the vector meson dominance model. 
The in-medium $B$ and $B^*$ meson masses necessary to evaluate the 
$\Upsilon$ and $\eta_b$ self-energies in symmetric nuclear matter, are calculated by 
the quark-meson coupling model.
In considering the unexpectedly larger contribution 
from the heavier vector meson $B^*B^*$ meson loop contribution, 
and the similar fact for the $J/\Psi$ mass shift due to the $D^*D^*$ meson loop, 
we regard our prediction for the $\Upsilon$ mass shift 
as taking the minimum meson loop contribution, 
namely, that is estimated by taking only the $BB$ meson loop contribution,  
as was practiced similarly for the $J/\Psi$ mass shift taking  
only the $DD$ meson loop contribution.
Our prediction by this only $BB$-loop, gives the in-medium 
$\Upsilon$ mass shift 
that varies from -16 MeV to -22 MeV at the symmetric nuclear matter saturation density 
($\rho_0 = 0.15$ fm$^{-3}$) for the cutoff mass values in the range from 2000 MeV to 6000 MeV.
For the $\Upsilon$ meson produced in a large nucleus with a sufficiently low relative momentum 
to the nucleus, the mass shift obtained suggests that it may be possible to form the 
$\Upsilon$-nucleus bound states with the only-$BB$-loop-based   
mass shift (potential). 
The study of the possible $\Upsilon$-nucleus bound states requires further investigations.

A detailed analysis is also made for the $\Upsilon$ self-energy meson loops 
due to the total ($BB+BB^*+B^*B^*$) contribution and the decomposition 
by comparing with the corresponding ($DD+DD^*+D^*D^*$) contribution and the decomposition  
for the $J/\Psi$ mass shift,  
focusing on the form factors in the interaction vertices   
using the correspondence between ($\Upsilon$ and $J/\Psi$), 
($B$ and $D$), and ($B^*$ and $D^*$) mesons. 
We have confirmed that, in the both cases of the $\Upsilon$ and $J/\Psi$ mass shifts, 
the heavier $B^*B^*$ and $D^*D^*$ 
meson loop contributions for the respective self-energies are larger than 
those of the corresponding lighter mesons, ($BB$ and $BB^*$) 
and ($DD$ and $DD^*$) meson loops, respectively. 
This fact suggests that our treatment for the vertices involving $B^*B^*$ mesons 
for the $\Upsilon$ self-energy, as well as the $D^*D^*$ mesons for the $J/\Psi$ self-energy,  
should be improved in treating the short distance fluctuations better.
Furthermore, we have chosen the same coupling constant value for 
$\Upsilon BB$, $\Upsilon BB^{*}$ and $\Upsilon B^{*}B^{*}$. 
A more dedicated study on this will be carried out in the near future.

Based on the detailed analysis on the $\Upsilon$ mass shift, 
we have also studied the $\eta_b$ mass shift on the same footing 
as that for the $\Upsilon$, based on an SU(5) effective Lagrangian. 
By this we have included only the $BB^*$ meson loop contribution for the $\eta_b$ 
self-energy as our prediction.
The obtained $\eta_b$ mass shift at symmetric nuclear matter saturation density 
ranges from -75 to -82 MeV for the same ranges of the cutoff mass values  
used for the $\Upsilon$ mass shift, from 2000 MeV to 6000 MeV.
For the $\eta_b B B^*$ coupling constant, we have used the SU(5) 
universal coupling constant determined by the $\Upsilon BB$ coupling constant   
by the vector meson dominance model with the experimental data.

We have also studied the $\Upsilon$ and $\eta_b$ mass shifts in a 
heavy quark (heavy meson) symmetry limit by calculating 
their mass shifts using the same coupling constant value   
with that for the corresponding $J/\Psi$ and $\eta_c$ mass shifts. 
For the $\eta_b$ mass shift, also a broken SU(5) symmetry from the $\Upsilon$ case 
has been studied within this limit. 
Our predictions for these cases at nuclear matter saturation density are, 
-6 to -9 MeV for $\Upsilon$, -31 to -38 MeV for $\eta_b$, 
and -8 to -11 MeV for $\eta_b$ with a broken SU(5) symmetry, where 
the corresponding mass shifts in the charm sector ones are, 
-5 to -21 for $J/\Psi$, -49 to -87 for $\eta_c$, and 
-17 to -51 for $\eta_c$ with a broken SU(4) symmetry. 
Thus, the bottomonium mass shifts are generally smaller than 
those of the corresponding charm sector in this limit, and the dependence on 
the cutoff mass value in the form factor is also smaller.
To see whether these or the other cases are closely realized in nature, 
further experiments are needed to get more information on the bottomonium-nucleon 
(bottomonium-(nuclear matter)) as well as those for the charmonium.

For all cases of the predicted mass shifts for the $\Upsilon$ and $\eta_b$ mesons, 
the variations in the predicted values for a wide range of the cutoff mass values 
(from 2000 to 6000 MeV) in the corresponding form factors, are small --- less than 10 MeV, 
and this fact reduces some ambiguity in the predictions 
originating from the cutoff mass values.

In addition, we have also performed an initial study for the effects of a form factor 
on the lowest order $\Upsilon$ and $\eta_b$ mass shifts --- our predictions. 
The different form factor applied gives a clearer physics picture 
for the interaction ranges between the $\Upsilon$-$B$, $\eta_b$-$B$ and $\eta_b$-$B^*$. 
Using the cutoff mass values based on the physics picture of the form factor, 
the calculated $\Upsilon$ and $\eta_b$ mass shifts have turned out to give  
similar values with those for the predicted values of   
the $\Upsilon$ and $\eta_b$ mass shifts obtained using the original form factors.

In the future we plan to perform an elaborated study on the form factors appearing 
in the $\Upsilon$ and $\eta_b$ self-energy vertices, as well as 
those corresponding in the $J/\Psi$ and $\eta_c$. 
Furthermore, we plan to study the $\Upsilon$-nucleus and 
$\eta_b$-nucleus bound states, and the effect of the meson widths.

\section{Acknowledgements}
\label{ack}
\noindent
GNZ was supported in part by the 
Coordena\c{c}\~ao de Aperfeiçoamento de Pessoal de N\'ivel Superior 
- Brazil (CAPES), and KT was supported by the Conselho Nacional de Desenvolvimento
Cient\'{i}fico e Tecnol\'{o}gico (CNPq)
Process, No.~313063/2018-4, and No.~426150/2018-0,
and Funda\c{c}\~{a}o de Amparo \`{a} Pesquisa do Estado
de S\~{a}o Paulo (FAPESP) Process, No.~2019/00763-0,
and this work was also part of the projects, Instituto Nacional de Ci\^{e}ncia e
Tecnologia --- Nuclear Physics and Applications (INCT-FNA), Brazil,
Process. No.~464898/2014-5, and FAPESP Tem\'{a}tico, Brazil, Process,
No.~2017/05660-0.



\end{document}